\documentstyle[12pt]{article}
\newcommand{\sect}[1]{\setcounter{equation}{0}\section{#1}}
\renewcommand{\theequation}{\arabic{section}.\arabic{equation}}

\begin{document}

\title{SEMI-CLASSICAL STABILITY OF SUPERGRAVITY VACUA}

\author{Marika M. Taylor-Robinson
\\DAMTP\\Silver Street\\Cambridge\\CB3 9EW}

\maketitle

\hfil{\bf Abstract\,\,\,\,\,\,}\hfil

\vbox{
\bigskip
\noindent

We discuss the existence of instantonic decay modes which would indicate
a semi-classical instability of the vacua of 
ten and eleven dimensional supergravity theories. Decay modes whose
spin structures are incompatible with those of supersymmetric vacua have
previously been constructed, and we present generalisations including
those involving non trivial dilaton and antisymmetric tensor fields. 
We then show that the requirement that any 
instanton describing supersymmetric vacuum decay
should admit both a zero momentum hypersurface from which we
describe the subsequent Lorentzian evolution and a spin structure at
infinity compatible with the putative vacuum excludes all such decay
modes, except those with unphysical energy momentum tensors which violate
the dominant energy condition.

}

\vfil
\noindent

\renewcommand{\thepage}{}
\pagebreak

\renewcommand{\thepage}{\arabic{page}}
\setcounter{page}{1}

\sect{Introduction}
\noindent

Supergravity theories exist in all spacetime dimensions $d$ with $d
\leq 11$, and are currently regarded as effective field theories of
superstring (and M) theories in some appropriate limit. 
Classical solutions of
the theories can be found by setting to zero the fermionic fields
together with their supersymmetric variations. We look for a vacuum in which
the space-time is of the form $B^4 \times K$ where $B^4$
is a maximally symmetric four-dimensional space (de Sitter space, anti
de Sitter space or Minkowski space) and K is a compact 
manifold; such a solution is consistent with the low-energy field
equations, with the dilaton field constant and all other fields
vanishing.
The conditions for finding supersymmetric generators that
leave the vacuum invariant restrict $B^4$ to be flat Minkowski space
and $K$ to be a manifold that admits at least one covariantly constant
spinor field.  This in turn constrains the possible holonomy groups of
$K$; for ten dimensional theories, $K$ must have a holonomy 
contained in $SU(3)$ \cite{Can}, implying that $K$ must
have a covering space that is $T^6$, $T^2 \times K3$ or a Calabi-Yau
space $K_{SU(3)}$. Similarly, for eleven dimensional supergravity, 
the holonomy of
$K$ is contained in $Spin(7)$, and $K$ has a covering space that is
$T^7$, $T^3 \times K3$, $K_{SU(3)} \times S^1$ or $K_{Spin(7)}$ (where the
latter is a manifold with the exceptional holonomy group 
$Spin(7)$) \cite{Du}.

It is important to have some criteria for determining whether $M^4
\times K$ is a reasonable candidate as the ground state of supergravity
theories, but our incomplete understanding of string (and M) theory dynamics 
makes this question difficult to answer in full.
The constraints above restrict the vacuum state to be Ricci-flat, with
the requisite holonomy; we must, however, also show that the
spectrum of the vacuum is stable and that there are no instantonic
decay modes, ie. we must thus impose
the conditions that $M^4 \times K$ should be stable at the classical
and semi-classical level, which leads to non-trivial conditions on the
vacuum manifold. 

 The first test of the stability of a space is to ask whether the
space is stable classically against small oscillations. Small
oscillations around $M^4 \times K$ will consist of a spectrum of massless
states (the graviton, gauge fields, dilaton etc) and an infinite
number of charged massive modes. The massless spectrum of the
heterotic string theory, which is the theory that we will consider
principally here, has been extensively discussed (see for
example \cite{Can}, \cite{String}, \cite{Gr} and \cite{Du}); there are
no exponentially growing modes with imaginary frequencies. The same
applies to that of eleven dimensional supergravity. 

 Even if a state is stable against small oscillations, it may be
unstable at the semiclassical level. This can occur if it is
separated by only a finite barrier from a more stable state; it will
then be unstable against decay by semiclassical barrier
penetration. To look for a semiclassical instability of a putative
vacuum state, one looks for a bounce solution of the classical
Euclidean field equations; this is a solution which asymptotically at
infinity approaches the putative vacuum state. If the solution is
unstable, then the Gaussian integral around that solution gives an
imaginary part to the energy of the vacuum state, indicating an
instability.

 The stability of Minkowski space at the semi-classical level 
as the unique vacuum state of general relativity was
proved by the positive energy theorem of Schoen and Yau \cite{ScYa}.
A completely different proof involving spinors satisfying a Dirac type
equation on a
three-dimensional initial value hypersurface was given by Witten
\cite{Wit1}, and shortly after re-expressed in terms of the Nester tensor 
\cite{Nes}.  Witten later
demonstrated the instability of the $M^4 \times S^1$ vacuum of
Kaluza-Klein theory \cite {Wit2}; the effective four-dimensional
vacuum decays into an expanding bubble of ``nothing''. However, this
decay mode is excluded by the existence of (massless) elementary
fermions. 

Instabilities of non-supersymmetric vacua of string theories were
discussed by Brill and Horowitz \cite{BH} who demonstrated 
that superstring theories admit instantonic decay modes
that asymptotically resemble toroidal compactifications
with constant gauge fields (but are incompatible with massless fermions).
Mazur \cite{M} showed that toroidal compactifications of multidimensional
Minkowski space-time are semiclassically unstable due to topology
change of the initial data hypersurface, and presented Euclidean
Schwarzschild p-branes as possible instantons corresponding to
tunnelling between different topologies. However, the instantons
discussed by Mazur do not correspond to vacuum instabilities 
since they do not take account of the incompatible spinor structures of
the instantons and (supersymmetric) vacua. 
Banks and Dixon \cite{BaDi} used conformal field theory
arguments to show that spacetime supersymmetry cannot be continuously
broken within a family of classical vacua and that two supersymmetric
vacua are infinitely far away. 
It was suggested in \cite{CGR} that if one takes into account 
target space duality all topology changing instabilities of 
toroidal vacua are impossible in the context of string theory.

More recently, another possible decay
mode of the Kaluza-Klein vacua has been constructed by Dowker et al. 
\cite{Do1},
\cite{Do2}. ``Magnetic'' vacua in Kaluza-Klein theory - vacua
corresponding to static magnetic flux tubes in four dimensions - may
decay by pair creation of Kaluza-Klein monopoles, though at a much
smaller rate than for decay by bubble formation; the pair creation 
decay mode is however
consistent with the existence of elementary fermions. 

\bigskip

In this paper, we investigate further possible decay modes of the
vacua of supergravity theories. Such instantons certainly do not
preserve the supersymmetry of the vacuum; however, we cannot {\it a priori}
exclude the possibility of the vacuum decaying into a state which
asymptotically admits the supersymmetry generators of the vacuum. That
is, there may exist solutions to the Euclidean field equations both
whose geometry is asymptotic to that of the background vacuum state,
and whose spin structure at infinity is compatible with that of the 
supersymmetric vacuum. 

In \S2 we describe the vacua of ten-dimensional supergravity theories,
and discuss new examples of twisted compactifications which give rise
to magnetic vacua in four dimensions. In the following section, we
discuss Ricci-flat instantons which describe decay of toroidal vacua
by bubble formation and pair creation of monopoles. 
In \S4 we consider more general instantons, relaxing the assumption
of Ricci flatness, and construct from five-dimensional charged black 
hole solutions decay modes involving non-vanishing dilaton and 
antisymmetric tensor fields. 

In \S5 we use extremal black hole solutions with non-degenerate
horizons to describe decay modes whose topology is not inconsistent
with asymptotically covariantly constant spinors, but whose energy
momentum tensors are unphysical, violating the dominant energy
condition. 

In \S6 we discuss more generally the existence of instantons
describing the decay modes of the supersymmetric vacuum; we consider
the formulation of Witten's proof of the positive energy theorem, and
show how this proof excludes the existence of physical decay modes
of a supersymmetric vacuum. In \S7 and \S8 we extend the discussion to
the Calabi-Yau vacuum of ten dimensional supergravity and to the vacua
of eleven-dimensional supergravity. Finally, in \S9 we present our
conclusions. 

Note that contrary to globally supersymmetric Yang-Mills theories,
supergravity is not renormalisable. This puts the entire subject of
instanton calculus in supergravity on a rather shaky basis; if however
we regard supergravity theories as low energy limits of superstring
theories, which are not expected to suffer from these deficiencies, to
the order to which supergravity theories are formally renormalisable
results from non-perturbative instanton calculations may be considered
as the limiting values of the corresponding exact string theory
results \cite{Ko}. 

\bigskip

Since manifolds of many different dimensions will abound, we will
adhere to the following conventions. The indices $M,N = 0,..,9$; $m,n
= 1,..,9$; $\mu,\nu = 0,..,3$; $i,j = 4,..,9$; $\alpha,\beta = 1,..,3$;
$a,b = 0,..4$; $I,J = 5,..,9$; $A,B = 0,..,5$, $F,G = 6,..,9$, $w,x =
0, 10$ and $f=1,..,16$. We use the mostly positive convention for 
Lorentzian metrics and $G$ will
denote metrics in the string frame whilst $g$ denotes metrics in the
Einstein frame. $c$ denotes induced metrics on boundaries at spatial infinity,
whilst $\hat{g}$ denotes induced metrics on spacelike hypersurfaces.
Hatted indices refer to an orthonormal frame whilst
unhatted indices refer to coordinate indices. $S$ denotes the Lorentzian
action and $S_{E}$ denotes the Euclidean action. $G_{d}$ refers to the
Newton constant in $d$ dimensions. 

\sect{Vacua of supergravity theories}
\noindent

Our starting point is the $d=10, N=1$ action that arises as a low
energy effective field theory from the heterotic string. 
We shall consider heterotic string theory for definiteness
but most of the discussion depends
only on the common sector of the low energy supergravity theory. 
For the massless bosonic fields of the theory (graviton ${G}_{MN}$, dilaton
${\Phi}$, antisymmetric tensor ${B}_{MN}$, 16 vector bosons 
${A}^{f}_M$) the action (in the string frame) takes the form:  
\begin{equation}
S =  \frac{1}{16 \pi G_{10}} \int d^{10}x \sqrt{-G} e^{-\Phi}
 \lbrace R(G) + (\partial\Phi)^2 
- \frac{1}{12} H^2 - \frac{1}{4} \rm{Tr}(F^2) \rbrace
\label{boselagrangian}
\end{equation}
As in general relativity, this action will give the correct field
equations but in calculating the action we must also include the
surface terms required to ensure unitarity. The corresponding action
in the Einstein frame $g_{MN} = e^{-\frac{\Phi}{4}} G_{MN}$ is:
\begin{equation}
S = \frac{1}{16\pi G_{10}} \int d^{10}x \sqrt{-g} \lbrace R 
- \frac{1}{8}(\partial\Phi)^2 - \frac{1}{12}e^{-\frac{\Phi}{2}} H^2 
- \frac{1}{4} e^{-\frac{\Phi}{4}} Tr(F^2) \rbrace \label{einfr}
\end{equation}
To find classical solutions of the supersymmetric theory,
we set to zero the fermion fields (gravitino
$\psi_M$, dilatino $\lambda$, gaugino $\chi$) 
together with their supersymmetric variations.
Assuming that all fields vanish except the graviton (and constant dilaton)
the conditions for finding a supersymmetric generator $\eta$ that
leaves the vacuum invariant reduce to: 
\begin{equation}
\delta\psi_M = D_M\eta =0
\label{spin}
\end{equation}
As is well-known, and was first shown in \cite{Can} (see also 
\cite{String}), for a vacuum 
state of the form $B^4 \times K$, where $B^4$ is a 
maximally symmetric four-dimensional space and $K$ is a compact 
six manifold, equation (\ref{spin}) implies that the maximally 
symmetric manifold must be flat Minkowski space, and $K$ must be a
Ricci-flat compact six manifold, which admits at least one covariantly
constant spinor of each chirality. The holonomy group of $K$ is thus
constrained to be a subgroup of the generic holonomy group of a
six-dimensional manifold $SO(6)$, and hence the covering space of $K$
must be $T^6$, $T^2 \times K3$ or a Calabi-Yau space. 

\bigskip

The simplest manifold which satisfies (\ref{spin}) and is Ricci-flat
is the flat torus; we may also consider another class of toroidal
vacua which asymptotically tend to static magnetic configurations in
four dimensions. Although these solutions are non-trivial four
dimensional configurations they are simply obtained from dimensional
reduction of Euclidean space with twisted identifications.
The construction of a static cylindrically symmetric flux tube in four
dimensions by dimensional reduction of five dimensional Minkowski
space in which points have been identified in a nonstandard way was
discussed in \cite{DGGH}, whilst more recently the construction was
generalised to obtain sets of orthogonal fluxbranes in higher
dimensional spacetimes \cite{Do2}. 

We will now consider a four dimensional vacuum solution in which there
are $p$ magnetic fields, arising from the ten-dimensional metric,
associated with $p$ distinct $U(1)$ isometry groups. 
We start with $10$-dimensional Minkowskian space and identify points
under combined spatial translations and rotations, ie.
\begin{equation}
ds^2 = - dt^2 + dz^2 + d\varrho^2 + \varrho^2 d\psi^2 +
\sum_{i}(dx^i)^2 \label{sp}
\end{equation}
where we identify points $(x^i,t,z,\varrho,\psi) \sim (x^i + 2\pi
\mu^i n^i,t,z,\varrho,\psi + \sum_{i}2\pi B^i \mu^i n^i + 2\pi n)$ and
$n,n^i$ are integers. We will usually assume that the $\mu^i$ are identical.
Since $\psi$ is already periodic, changing $\sum_{i} \mu^i B^i$ 
by an integer does not change the identifications; thus inequivalent
spacetimes are obtained only for $-1/2 < \sum_{i}\mu^i B^i \leq
1/2$. Changing each $B^i$ by a multiple of $1/\mu^i$ leads to
equivalent spacetimes in $(4+p)$ dimensions, though the four
dimensional configurations are not equivalent. Geometrically. 
this spacetime is obtained by starting
with (\ref{sp}) and identifying points along the closed orbits of the
Killing vectors $l^i = \partial_{x^i} + B^i \partial_{\psi}$.  
Introducing a new
coordinate $\bar\psi = \psi + \sum _{i}B^i x^i$, we may rewrite the metric as
\begin{equation}
ds^2 = -dt^2 + dz^2 +  d\varrho^2 + (dx^i)^2 + \varrho^2
(d\bar\psi + \sum_{i} B^i dx^i)^2 \label{orr}
\label{mgf}
\end{equation}
Dimensionally reducing along the $6$ Killing vectors
$k^i= \partial_{x^i}$, the four-dimensional
metric in the Einstein frame $g_{\mu\nu}$ is related to the
$10$-dimensional metric by:
\begin{equation}
{g}_{MN}=\left ( \matrix{e^{\varphi}g_{\mu\nu} +
 \sum_{i,j}  \xi_{ij}A^{i}_{\mu}A^{j}_{\nu} 
& \sum_{i} A^{i}_{\mu} \xi_{ij} \cr
  \sum_{j} A^{j}_{\nu}\xi_{ij} & \xi_{ij}} \right)
\end{equation} 
with Kaluza Klein gauge fields $A^{i}$ and $\varphi$ the four-dimensional
dilaton defined by $\varphi = \Phi - \frac{1}{2}
\rm{ln} (det \xi_{ij})$. From (\ref{orr}), we find that:
\begin{equation}
\xi_{ii} = ( 1 + (B^i)^2 \varrho^2)
\end{equation}
\begin{equation}
\xi_{ij} = B^{i}B^{j} \varrho^2 
\end{equation}
\begin{equation}
A^{i}_{\bar\psi} = \frac{B^i \varrho^2 [1 - 2 \sum_{j \neq
    i}B^{j}A^{j}_{\bar\psi}]} {(1 + (B^i)^2 \varrho^2)}
\label{fields}
\end{equation}
The off-diagonal terms in the internal metric imply that the torus is
not a direct product of circles. The four-dimensional metric is given by 
\begin{equation}
ds_{ein}^2 = (1 + \sum_{i} (B^i)^2 \varrho^2)^{\frac{1}{2}}
[-dt^2 + dz^2 + d\varrho^2] +
g_{\bar{\psi} \bar{\psi}}d\bar{\psi}^2
\label{solut}
\end{equation}
where $g_{\bar{\psi}\bar{\psi}}$ is defined by: 
\begin{equation}
g_{\bar{\psi}\bar{\psi}} = (1 + \sum_{i} (B^i)^2 \varrho^2)^{\frac{1}{2}}
[\varrho^{2}-\sum_{i,j} \xi_{ij} A^{i}_{\bar{\psi}} A^{j}_{\bar{\psi}}]
\end{equation}  
The gauge fields in four dimensions are obtained by solving the $p$
simultaneous equations defined in (\ref{fields}); in the limit that
only 
one field is non-zero, the solution reduces to the static magnetic
flux tube. Asymptotically, each gauge field $A^i \rightarrow
\frac{1}{pB^i}$; the gauge fields correspond to magnetic
fields which are uniform at infinity. 

We may also dimensionally reduce along the Killing vectors
$\tilde{k^i} = k^i + (n^i/\mu^i)\partial_{\bar\psi}$; the corresponding
four dimensional solution is unchanged, except that the magnetic field
is modified to $\tilde{B^i} = B^i
+ n^i/\mu^i$, and in this way all values of the four-dimensional magnetic
fields associated with each $U(1)$ can be obtained. 
For every $B^i \neq 0$, the proper length of the circles
in the $i$th direction grows linearly with $\varrho$ for large $\varrho$;
thus, we can view the solution as an approximation to physical fields
which is valid only for $\varrho \ll 1/ B^i$, in which range the three
dimensional space is approximately flat, and the internal circles have
approximately constant length. In order for the internal directions to
remain unobservable, we must consider length scales which are large
compared to their size: $\varrho \gg \mu^i$. The two restrictions imply 
a limited range of applicability of the spacetime, $B^i \ll 1/\mu^i$, 
which can include
large magnetic fields only if the compactified dimensions are of the
Planck scale. Since the different dimensional reductions change $B^i$
by multiples of $1 /\mu^i$, for given $\mu^i$, at most one is
physically reasonable. 

\bigskip

These solutions can be obtained by the action of generating transformations on
the original Kaluza-Klein solution of \cite{DGGH}; the required
transformations are an $O(6)$
subgroup of the $O(6,22)$ T-duality group of the 
four-dimensional theory. The transformation acts as:
\begin{eqnarray}
\tilde{A}^{i}_{\mu} &=& \Omega_{ij}A^{j}_{\mu} \nonumber \\
\tilde{\xi} &=& \Omega^{T} \xi \Omega 
\end{eqnarray}
where $\Omega$ is an $O(6)$ invariant matrix, and all other fields
are left invariant. The particular transformation required here
is (assuming that the radii of the compactified directions are identical)
$R_{6}(\vec{k})$, a six-dimensional rotation that
rotates an arbitrary six-dimensional vector in direction $\vec{k}$ 
into a vector of the same magnitude with only one component non-zero. 

\bigskip

Consider a solution for which $p$ fields are non-zero and equal
to $B$:
\begin{eqnarray}
ds_{ein}^2 &=& e^{-\varphi}[-dt^2 + dz^2 + d\varrho^2] + \varrho^2 
e^{\varphi}d\bar\psi^2 \nonumber \\
A^{i}_{\bar\psi} &=& \frac{B \varrho^2}{(1 + pB^2\varrho^2)} \\
e^{-2\varphi} &=& (1 + p B^2 \varrho^2)
\end{eqnarray}
This has the interpretation of a flux tube along the z-axis,
associated with $p$ magnetic fields, and is the required background
for nucleation of monopoles carrying charges with respect to p $U(1)$
fields.  

Even though these spacetimes are locally flat, the nontrivial
identifications imply that if a vector is parallely transported around
each $S^1$, it will return rotated by an angle $2 \pi \mu^i B^i$. It
follows that for one spin structure, parallel propagation of a spinor
around the $i$th direction results in the spinor acquiring a phase
$e^{\pi \mu^i B^i \gamma}$, where $\gamma$ is a
generator of the Lie algebra of $SO(9,1)$ (spinor representation). 
For the other spin structure, parallel propagation gives a
phase $-e^{\pi \mu^i B^i \gamma}$. For small $B^i$, the natural 
generalization of the standard choice of spinor structure for a
supersymmetric vacuum is the first choice. 
The magnetic vacua evidently admit no covariantly constant spinors,
whereas for the standard metric on the torus, $32$ constant spinors 
are admitted.

\bigskip

Note that the invariance of the low energy effective action under
the $O(6,22)$ T-duality group, and the
invariance of the equations of motion under $SL(2,R)$ S-duality transformations
allows us to generate further solutions. 
We apply a particular S-duality transformation $\varphi \rightarrow
\tilde{\varphi} = -\varphi$ (corresponding to strong/weak coupling
interchange), $F^{i+6}_{\mu\nu} \rightarrow
\tilde{F}^{i+6}_{\mu\nu} = e^{-2\varphi}\xi_{ij}\bar{F}^{j}_{\mu\nu}$ 
(with $\bar{F}$ the dual of $F$) and  
$F^{i}_{\mu\nu} \rightarrow \tilde{F}^{i}_{\mu\nu} = 0$. Then,  
rescaling to the string
metric, $G_{\mu\nu} = e^{\tilde{\varphi}}g_{\mu\nu}$, the
four-dimensional solution becomes:
\begin{eqnarray}
ds_{str}^2 &=& e^{2\tilde{\varphi}}(-dt^2 + dz^2 + d\varrho^2) +
\varrho^2d\bar{\psi}^2 \nonumber \\
\tilde{F}^{i+6}_{\tau z} &=& \frac{2B}{(1+B^2\varrho^{2}p)^{\frac{1}{2}}}
\\
e^{2\tilde{\varphi}} &=& (1 + B^2\varrho^2 p)
\end{eqnarray} 
Thus, the only non-vanishing gauge fields are those originating from
the off-diagonal components $B_{\mu i}$ of the two-form in ten dimensions.
The solution describes an `electric' flux tube, associated with
$p$ gauge fields; each gauge field asymptotically approaches zero. 

By applying a general $O(6,22)$ generating transformation, we can obtain
four-dimensional solutions describing tubes of magnetic flux
associated with the $U(1)^{28}$ gauge group of the heterotic
theory; these are the required backgrounds for
nucleation of other topological defects, such as H-monopoles. 

\section{Ricci-flat instantons}
\noindent

We firstly consider instanton solutions of the Euclidean field
equations in ten 
dimensions, in which all fields except the graviton and the (constant)
dilaton vanish, implying that $R_{MN} = 0$. The asymptotic geometry of
the instanton is $R^4 \times T^6$; we defer the discussion of
Calabi-Yau vacua to \S7. 
Evidently decay modes cannot preserve all the supersymmetry; 32
constant spinors requires trivial holonomy, implying that the
solution admits a flat metric. However, if an instanton is to describe
a possible vacuum decay mode, it must asymptotically admit the
constant spinors of the background.
Vacuum instability - which in many cases will correspond to physical 
formation processes - will hence
result only from considering non-supersymmetric
states which are not simple metric products, but rather contain topological
defects such as monopoles or p-branes. 

\bigskip

The instantons will usually globally admit a $U(1)^6$ 
isometry group, as well as a hypersurface orthogonal Killing
vector which we use to Wick rotate the solution to describe the
subsequent Lorentzian evolution. We may also of course consider instantons
which admit such $U(1)$ isometries only asymptotically, although it is
unclear how the effective four-dimensional solution can be interpreted
in this case. Fixed points 
of these isometries will lead to apparent singularities such as bubbles and
monopoles in the lower-dimensional spacetime; the structure of these 
fixed point sets determines whether the instanton has a spin structure
consistent with that of the background. 

For a ten-dimensional instanton, the fixed point set must have
dimension $10$, $8$, $6$, $4$, $2$, $0$; the classification of four-dimensional
gravitational instantons in terms of the fixed points sets of a $U(1)$
isometry was discussed in \cite{GH} and reviewed in \cite{EGH}.
This work has been generalised to
higher dimensions in \cite{Do2} and \cite{GPP}. 

If the isometry admits no fixed point sets, there is {\it a priori} no
obstruction to choosing the spin structure of the instanton to be
consistent with that of the background. 
If however the fixed point set of the isometry is eight-dimensional, spinors
must be antiperiodic about a closed orbit of the isometry at infinity, and 
the spin structure is incompatible with that of the (supersymmetric)
vacuum. The twisted boundary conditions break supersymmetry, and, although
this supersymmetry breaking can be made arbitrarily weak by taking the
compactified directions to be arbitrarily large, the action of the instanton
diverges as the radii approach infinity, implying that the rate of decay
of the vacuum goes to zero. 

The obvious example is the five-dimensional Euclidean Schwarzschild
solution crossed with a flat torus (a decay mode first considered in
\cite{Wit1}):
\begin{equation}
ds^2 =  dx^{I}dx_{I} + (1 - \frac{\mu}{r^2})d\tau^2 + 
\frac{dr^2}{(1 - \frac{\mu}{r^2})} + r^2 d\Omega_{3}^{2} 
\label{sl}
\end{equation}
where the periodicity of $\tau$ is $\Delta\tau = 2\pi \sqrt{\mu}$ and 
the range of $r$ is $r \geq \sqrt{\mu}$. Since the topology of 
the solution is 
$R^2 \times S^3 \times T^5$, the spin structure is incompatible with 
that of the supersymmetric vacuum. The action for this instanton is
obtained from the boundary term:
\begin{equation}
S_{E} = - \frac{1}{16\pi G_{10}} \oint d^{9}x \sqrt{c} \lbrace {\cal K} -
{\cal K}_{0} \rbrace
\end{equation}
with ${\cal K}$ the trace of the second fundamental form of the boundary at
infinity, and ${\cal K}_{0}$ the corresponding term in the background. The
action is hence:
\begin{equation}
S_{E} = \frac{\pi \mu}{8 G_{4}} \label{tmi}
\end{equation}
and thus the rate of vacuum decay does indeed vanish 
as the radii of the compactified
directions increase, that is, as the supersymmetry breaking becomes 
arbitrarily weak. Furthermore, the decay rate is small if $\sqrt{\mu}
\gg$ Planck length, and it is only in this case that the semiclassical
calculation is reliable. 

\bigskip

If the fixed point set of the isometry has dimension less than eight, the
spinors need not necessarily
be antiperiodic about a closed orbit of the isometry at infinity. As an
example, we may consider dimensional reduction of (\ref{sl})
along the Killing vector $k = \partial_{\tau} +
\frac{1}{R}\partial_{\psi} $  \cite{Do1} (where $R$ is the radius of
the circle at infinity) which 
describes monopole pair creation in a background field.
It is possible to choose the spin structure to be consistent at
infinity with that of the solution describing the background field
(\ref{mgf}), but the magnitude of the magnetic field lies far outside 
the physical range of validity.

\bigskip

As also discussed in \cite{Do1}, we may consider instantons
which are a product of
$T^5$ and the five-dimensional Euclidean Kerr-Myers-Perry solution:
\begin{eqnarray}
ds^{2} &=& (dx^{F})^2 + dx^{2} + dy^{2} + {\rm{sin}}^{2}\theta(r^2 - 
\alpha^{2})(d\psi)^{2} + \frac{\rho^{2}}{r^2 - \alpha^{2} - \mu} dr^2
\nonumber \\
&& + \rho^{2}d\theta^{2} + r^2 {\rm{cos}}^2\theta d\tau^2  
 - \frac{\mu}{\rho^{2}}(dx + \alpha \sin^{2}\theta (d\psi))^{2} \label{q1}
\end{eqnarray} 
where the index $F$ runs over the coordinates of the remaining $T^4$
and $\rho^2 = r^2 - \alpha^2 \rm{cos}^{2}\theta$. The most general
such solution is labelled by one mass parameter and two angular momentum
parameters (associated with the $(S0(2))^2 \times O(2)$ isometry
group), but for simplicity we take only the mass parameter $\mu$ and
one angular momentum parameter $\alpha$ to be non-zero. 

Reduction along $\partial_{x} + \frac{\alpha}{\mu}\partial_{\psi}$,
which has an eight-dimensional fixed point set, leads to decay of the
four-dimensional vacuum by bubble formation, whilst reduction along
$\partial_{x} + (\frac{\alpha}{\mu} + \frac{1}{R})\partial_{\psi}$,
which has a six-dimensional fixed point
set, leads to decay by monopole pair production. In the latter case,
the four-dimensional field $B = \frac{\alpha}{\mu} + \frac{1}{R}$, and
for $\frac{\alpha}{\mu} \sim -\frac{1}{R}$, we obtain fields of physical
validity. The pair creation decay mode has a spin structure
consistent with that of the magnetic vacuum and the action for the 
instanton (\ref{q1}) is:
\begin{equation}
S_{E} = \frac {\pi R^{2}}{8 G_{4} (1 - R^2 (\frac{\alpha}{\mu})^{2})} 
\label{occ}
\end{equation}
so for physical $B \ll 1/R$ the decay rate $\Gamma \sim e^{-S_{E}}$ is
very small. 

\bigskip

In \S 2, we showed that by applying a generating transformation to
a solution in which
there was a single non-zero magnetic field in four dimensions we could
obtain a solution in which there were several non-zero (metric) $U(1)$
fields in four dimensions. Using the same generating techniques here, we
expect to obtain more general solutions describing the pair creation
of monopoles carrying charge with respect to $p$ $U(1)$ gauge fields
in a background of $q$ $U(1)$ gauge fields ($p,q \leq 6$). 
Monopoles carrying several different
charges have recently been constructed \cite{FK} within heterotic 
string theory by the supersymmetric uplifting of four-dimensional
monopole solutions; we now discuss their nucleation. 

\bigskip

The most general solution Ricci-flat solutions
will be obtained from the three parameter five dimensional
Euclidean Kerr-Myers-Perry solution (constructed in \cite{Per}) 
crossed with a flat five torus by
first applying an $O(6)$ transformation to $\xi$ and $A$, and then 
identifying points along closed orbits of $\partial_{x^{i}} + B^{i}
\psi$ where the $B^{i}$ are chosen so that the action of the isometry is
periodic. 

\bigskip

We present an illustrative solution, describing pair creation of
monopoles carrying two identical $U(1)$ charges within two (equal) background
fields. We identify points along closed orbits of $\partial_{x} + B
\partial_{\psi}$, and $\partial_{y} + B \partial_{\psi}$ in (\ref{q1})
and then introduce $\bar{\psi} = \psi - B(x+y)$. 
The required generating transformation is:
\begin{equation}
\Omega = \left( \matrix{ R_{2} & 0  \cr
                            0  & I_{4}}
          \right) 
\end{equation}
where $\Omega$ acts on $\xi$ and $A$ as in \S 2.1, and
$R_{2}$ generates a two-dimensional transformation by $\pi/4$. 
The resulting solution is:
\begin{eqnarray}
ds^2 &=& (dx^{F})^2 + (1-\frac{\mu(1+\alpha \sqrt{2}B \rm{sin}^{2}\theta)^2)}{2
  \rho^2} + B^2 (r^2 - \alpha^2)\rm{sin}^{2}\theta)dx^2
\nonumber \\
&& \ + 2 \ (- \frac{\mu(1 + \alpha \sqrt{2} B\rm{sin}^{2}\theta)^2}{2\rho^2} + 
 B^2 (r^2 - \alpha^2)\rm{sin}^{2}\theta)dx dy \\
&& \ + \ (1 - \frac{\mu(1 + \alpha \sqrt{2} B \rm{sin}^{2}\theta)^2}{2
  \rho^2} + B^2 (r^2 - \alpha^2)\rm{sin}^{2}\theta)dy^2
\nonumber \\
&& \ + \ (B^2 (r^2 - \alpha^2)\rm{sin}^{2}\theta\sqrt{2} -
    \frac{\mu \alpha \rm{sin}^{2}(\theta)(1 + \alpha \sqrt{2} B
      \rm{sin}^{2}\theta)\sqrt{2}}{2 \rho^2}) dx d\bar{\psi} \nonumber \\
&& \ + \ ( B^2 (r^2 - \alpha^2)\rm{sin}^{2}\theta\sqrt{2} -
    \frac{\mu \alpha \rm{sin}^{2}(\theta)(1 + \alpha \sqrt{2} B
    \rm{sin}^{2}\theta)\sqrt{2}}{2 \rho^2}) dy d\bar{\psi} \nonumber \\
&& \ + \ \frac{\rho^2}{r^2 - \alpha^2 - \mu} dr^2 + \rho^2 d\theta^2 
     + r^2 \rm{cos}^2\theta d\tau^2 \nonumber \\
&& \ + \ ((r^2-\alpha^2)\rm{sin}^{2}\theta - \frac{\mu\alpha
  \rm{sin}^{4}\theta}{\rho^2}) d\bar{\psi}^2 \nonumber 
\end{eqnarray}
where $\rho^2 = r^2 - \alpha^2 \rm{cos}^{2}\theta$. 
The four-dimensional solution obtained by dimensionally reduced along
closed orbits of $\partial_{x}$ and $\partial_{y}$ is:
\begin{eqnarray}
ds_{4}^{2} &=& e^{-\varphi} \lbrace r^2 \rm{cos}^{2}\theta d\tau^2 + 
               \frac{\rho^2}{r^2 - \alpha^2 - \mu} dr^2  \\
  &&  + \rho^2 d\theta^2 + e^{2\varphi} 
(r^2 - \alpha^2 - \mu)\rm{sin}^{2}\theta)d\bar{\psi}^2 \rbrace \nonumber \\
e^{-2\varphi} &=& 1 - \frac{\mu}{\rho^2} (1 + \alpha \sqrt{2} B
\rm{sin}^{2}\theta)^2 + 2 B^2 (r^2 - \alpha^2) \rm{sin}^{2}\theta \nonumber \\ 
A^{x}_{\bar\psi} &=& \frac{\hat{g}_{yy}\hat{g}_{x\bar{\psi}} -
 \hat{g}_{xy}\hat{g}_{y\bar{\psi}}}{\xi}\nonumber \\
A^{y}_{\bar\psi} &=& \frac{\hat{g}_{xx}\hat{g}_{y\bar{\psi}} -
 \hat{g}_{xy}\hat{g}_{x\bar{\psi}}}{ \xi}\nonumber
\end{eqnarray}
where $\xi = e^{-2\varphi}$ is the determinant of the metric on the
torus; the ten-dimensional solution is complete and non-singular.

If we take $B = \alpha/\sqrt{2} \mu$, then the metric is
singular over all the horizon $r_{h}^2 = \alpha^2 - \mu$; the solution 
describes a generalised bubble decay mode of the vacuum. 

We may also take $B = \alpha/\sqrt{2}\mu + n/\sqrt{2}R$, in which case
$e^{-2\varphi}$ vanishes only at the poles of the horizon; this is the
solution describing pair creation of monopoles. The horizon is a line,
which is smooth provided that we take $n = \pm 1$; for $ \left | n
\right | > 1$, the singularities at the poles are joined by a string.
Since 
\begin{equation}
R = \frac{\mu}{\sqrt{\alpha^2 + \mu}} < \frac{\mu}{\alpha}
\end{equation}
to obtain four-dimensional magnetic fields 
of physical magnitude we need either $\alpha/\mu$ negative, close to 
$-1/R$ and $n=1$ or $\alpha/\mu$ positive, close to $1/R$ and $n=-1$. 
 
\bigskip

Consider the spin structure of the transformed solution. A spinor
parallely transported about an orbit of $l_{x} = \partial_{x} +
\alpha/\sqrt{2}\mu \partial_{\psi}$ can be shown to pick up a phase 
of $-e^{\pi R\sqrt{2} (\alpha/\sqrt{2}\mu) \gamma}$; the same phase is
picked up by a spinor transported about an orbit of $l_{y} = \partial_{y} +
\alpha/\sqrt{2}\mu \partial_{\psi}$. The four-dimensional magnetic
fields are $B^{x} = B^{y} = \alpha/\sqrt{2}\mu$, so this decay mode
by bubble nucleation is incompatible with the vacuum spin
+structure, defined by phases of 
$e^{\pi R\sqrt{2} (\alpha/\sqrt{2}\mu) \gamma}$.

If we take $B^{x} = B^{y} = \alpha/\sqrt{2}\mu + 1/\sqrt{2}R$ and 
dimensionally reduce along $l_{x}' = l_{x} + 1/\sqrt{2}R$ and 
$l_{y}' = l_{y} + 1/\sqrt{2}R$, then the phase change about an orbit
of $l'$ is found to be $-e^{\pi R (B\sqrt{2} - 1/R) \gamma} = e^{\pi R
  B \sqrt{2} \gamma}$ which is consistent with the vacuum.   

\bigskip

The action for this decay mode can be compared to (\ref{occ});
the transformation does not change the action, but after ensuring
that the unit of charge is the same in each case, we find that
\begin{equation}
S_{E}(1,1) = 2 S_{E}(1,0)
\end{equation}
where the notation specifies the charges carried by the
monopoles. Thence the rate of decay by creation of monopoles carrying $(1,1)$
charges is approximately half as large as the rate of decay by creation
of monopoles carrying $(1,0)$ charges (of the same magnitude); as we
would expect, the higher the charge, the smaller the rate of decay. 

\bigskip

By applying a more general $O(6,22)$ transformation to these solutions,
we might expect to obtain instantons describing the pair creation of
other types of monopoles, such as H monopoles, within the backgrounds
discussed in \S2. Although a large
class of solutions can be obtained by generating transformations, 
most of them will be singular and incomplete; the
nature of the ``dual'' geometry depends 
on the fixed points of the isometry with respect to which we dualise
and fixed points of the isometry in the original solution generically 
become singular points in the dual solution. 

For example, the solution appropriate to H-monopole nucleation is given by
the (Buscher) transformation (see \cite{GPR} for a review of T-duality
in string theory):
\begin{eqnarray}
g_{55} & \rightarrow & 1/g_{55} \nonumber \\
A^{1}_{\bar\psi} & \rightarrow & 0  \\
B_{5\bar\psi} & \rightarrow & A^{1}_{\bar{\psi}} 
\nonumber 
\label{Td}
\end{eqnarray}
with all other fields invariant. Dualisation with respect to the
isometry $\partial_{x}$ which has a fixed point set at $r=R$, $\theta = 0, \pi$
leads to a solution which is singular at these points (in both
string and Einstein frames); thus we cannot
interpret the solution as describing pair creation. 

\section{More general instantons}
\noindent

We have so far discussed only Ricci-flat instantons; evidently, more general
decay modes of the vacuum involving non-zero gauge, antisymmetric
tensor and dilaton fields should also be taken into
account. Consistency with the background requires that all fields
are asymptotically constant; 
these solutions were considered to some extent in \cite{BH} and we
suggest generalisations here. 

The decay modes presented in \S3 involved five-dimensional Euclidean
black hole solutions, with a non-trivial topology $R^{2} \times S^{3} 
\times T^{5}$ and in looking for generalisations it is natural to consider
electrically charged black hole solutions in five dimensions. 
\footnote{Five-dimensional black holes may carry a magnetic charge with respect
  to the three form field strength, but the latter takes the form $H =
  P \epsilon_{3}$, and does not asymptotically vanish, so we need not
  consider such solutions here.} 
In the following two
sections, we work with the effective five-dimensional action, and
implicitly take the product of the five-dimensional solution
with a flat torus.
Following the general prescription for dimensional reduction given in
\cite{Sen}, we obtain from (\ref{boselagrangian}) an
action in the string frame containing the terms:
\begin{equation}
S = \frac{1}{16 \pi G_{5}} \int d^{5}x \sqrt{-G} e^{-\phi}
\lbrace R(G) +  (\partial \phi)^{2} - \frac{1}{12} H^{2} -
\frac{1}{4}F^{2} \rbrace \label{act}
\end{equation}
with $F$ deriving from the left current algebra, and $G_{5} = G_{10}/V$ 
where $V$ is the volume of the $T^{5}$. Rescaling to the Einstein
frame $g_{ab} = e^{-2\phi/3}{G}_{ab}$, we obtain an action:
\begin{equation}
S = \frac{1}{16 \pi G_{5}} \int d^{5}x \sqrt{g} \lbrace R -
\frac{1}{3} (\partial\phi)^{2} - \frac{1}{12}e^{-\frac{4\phi}{3}}
H^{2} - \frac{1}{4} e^{-\frac{2\phi}{3}} F^{2} \rbrace
\end{equation}
We may now invoke Poincar\'{e} string-particle duality in five
dimensions to relate the three form field strength to its dual:
\begin{equation}
e^{-\phi} H^{abc} = \frac{1}{2! \sqrt{-G}} \epsilon^{abcde}
  \bar{F}_{de}
\end{equation}
which gives us the following action in terms of the axionic field
strength $\bar{F}$:
\begin{equation}
S = \frac{1}{16 \pi G_{5}} \int d^{5}x \sqrt{g} \lbrace R -
\frac{1}{3} (\partial\phi)^{2} - \frac{1}{4}e^{\frac{4\phi}{3}}
\bar{F}^{2} - \frac{1}{4} e^{-\frac{2\phi}{3}} F^{2} \rbrace \label{Act}
\end{equation} 
In \S5 we shall consider more general solutions with both of these 
gauge fields non-trivial but we begin with a
particularly simple five-dimensional (electrically) charged black hole
solution:
\begin{eqnarray}
ds^{2}_{str} &=& -
\frac{(1-\frac{\mu}{r^{2}})}{(1+\frac{k\mu}{r^2})^{2}}dt^{2}
+ \frac{dr^{2}}{(1-\frac{\mu}{r^{2}})} + r^2 d\Omega_{3}^{2} \nonumber \\
e^{-\phi} &=& (1+\frac{k\mu}{r^2}) \\
A &=& - \frac{1}{\sqrt{2}} \frac{\mu\rm{sinh}\delta}{(r^2+ k\mu)}
dt \nonumber
\end{eqnarray}
where $k = (\rm{cosh}(\delta)-1)/2$, and 
$A$ is the gauge potential associated with the field strength $F$. 
Such a solution was first constructed in \cite{GM}, and the Euclidean
section was discussed in \cite{BH}. We define the charge as:
\begin{equation}
Q_{f} = \frac{1}{16\pi} \int \ast F
\end{equation}
and, with this convention, $ Q_{f} = (\sqrt{2}/8)\pi\mu\rm{sinh}\delta$. 
We now look for a Euclidean section on which all the fields are real,
by rotating $t \rightarrow i \tau$. To obtain a real gauge potential
on the Euclidean section, we must also rotate $Q_{f} \rightarrow -i Q_{f}$
and hence $\rm{sinh}\delta \rightarrow -i (\rm{sin} \delta)$, giving the
solution:
\begin{eqnarray}
ds^{2}_{str} &=& \frac{(1-\frac{\mu}{r^{2}})}
{(1 - \frac{\kappa\mu}{r^{2}})^{2}} d\tau^{2} + \lbrace \frac{dr^{2}}{(1
  -\frac{\mu}{r^{2}} )} + r^{2} d\Omega_{3}^{2} \rbrace \\
A &=& - \frac{1}{\sqrt{2}} \frac{\rm{sin}\delta(r^2 - \mu)}
{(1- \kappa)(r^2 - \kappa\mu)} d\tau \nonumber
\end{eqnarray}
where $\kappa = (1-\rm{cos}\delta)/2$ and we have added a pure gauge
term to the potential so that $A^{2}$ is non singular at $r =
\sqrt{\mu}$. \footnote{Evidently the potential approaches a non-zero 
  constant at
  infinity, and thence a (purely gauge) Maxwell potential must exist
  in the background also; this presents no problem since
  the most general supersymmetric vacua may have constant gauge potentials.}
The coordinate $r$ is now restricted to 
$r \geq \sqrt{\mu}$ and we must identify $\tau$ with period
$ \Delta \tau = 2 \pi (1 - \kappa) \sqrt{\mu}$. The limit $\kappa = 0$
corresponds to an uncharged solution, whilst in the limit $\kappa =1$
the solution becomes singular. Since $0 \leq \kappa < 1$, the radius 
at infinity becomes smaller as $\kappa$ approaches its maximum value. 

\bigskip

By rotating one of the coordinates on the sphere, we obtain:
\begin{equation}
ds^{2}_{str} =  \frac{(1-\frac{\mu}{r^{2}})}
{(1 - \frac{\kappa\mu}{r^{2}})^{2}}
 d\tau^{2} + \frac{dr^{2}}{(1
  -\frac{\mu}{r^{2}} )} + r^{2} \rm{cosh}^{2}t (d\Omega_{2}^{2}) 
  - r^{2} dt^{2} 
\end{equation}
Since there are no terms of order $1/r$ in the fields, this solution has
zero mass and charge, which 
is consistent with the fact that it results from the decay of
a vacuum which certainly has zero mass and charge. 
Since the topology of the solution is $R^{2} \times S^{3}$, and the
Killing vector $\partial_{\tau}$ has a fixed point set of dimension 
three at $r = \sqrt{\mu}$, spinors must be antiperiodic about the
imaginary time direction, which prevents the solution from describing
the decay of a supersymmetric vacuum. 

It is straightforward to calculate the Euclidean action for this
solution; this is most easily done in the string frame, since we can
convert the volume term (\ref{act}) to a surface term \cite{BH} 
using the dilaton field equation:
\begin{equation}
S_{E} = -\frac{1}{8\pi G_{5}}  
\oint d^{4}x \sqrt{c} \lbrace (n \cdot 
\partial e^{-\phi}) + e^{- \phi} {\cal K} - e^{-\phi_{0}} {\cal K}_{0}
\rbrace \label{sa}
\end{equation}
where we now include the appropriate surface terms (and $\phi_{0}$ is
the asymptotic value of the dilaton field). The action is thus:
\begin{equation}
S_{E} = \frac{\pi\mu}{8 G_{4}} ( 6 \kappa + 1 ) 
\end{equation}
with $G_{5} = G_{4}/\Delta\tau$; this is consistent with
the Schwarzschild action given in \S3 when $\kappa = 0$ as required. 
Re-expressing this in terms of the radius at infinity:
\begin{equation}
S_{E} = \frac{\pi R^2}{8 G_{4}} \frac{(6 \kappa + 1)}{(1 - \kappa)^2}
\end{equation}
As before, the decay rate
$\Gamma \sim e^{-S_{E}}$ goes to zero as the supersymmetry breaking
becomes arbitrarily small, and, for given radius $R$, the vacuum
decay described by this solution is slower than decay via the
Ricci-flat solutions of \S3. If we let $\kappa \rightarrow 1$, with
the radius $R$ finite, the action diverges, and the radius of the 
``horizon'' approaches infinity. Letting $\kappa \rightarrow 1$
with $\mu$ constant gives a finite action, decreasing radius at infinity
and an unstable horizon (since the horizon is singular for $\kappa =1$).
In the limit of small charges, $Q_f \ll \mu$, 
it is straightforward to show that:
\begin{equation}
S_{E} = \frac{\pi\mu}{8G_{4}} ( 1 + \frac{48 Q_f^2}{\pi^2\mu^2})
\end{equation}
with the radius at infinity approximately $\mu$.
 
\bigskip

We obtain the effective four-dimensional solution using the procedure
given in \cite{Sen} as:
\begin{equation}
G_{ab} = \left( \matrix{ e^{\varphi}g_{\mu\nu} +
  G_{\tau\tau}A_{\mu}A_{\nu} & G_{\tau\tau}A_{\nu} \cr
  G_{\tau\tau} A_{\nu} & G_{\tau\tau}} \right)
\end{equation}
where $g_{\mu\nu}$ is the metric in the Einstein frame and
$\varphi = \phi - \frac{1}{2}\rm{ln}G_{\tau\tau}$. Then
the four-dimensional metric in the Einstein frame is:
\begin{eqnarray}
ds_{ein}^{2} &=& e^{-\varphi} \lbrace 
\frac{dr^{2}}{(1-\frac{\mu}{r^{2}})} + r^{2} \rm{cosh}^2 t
(d\Omega_{2}^{2}) - r^2 dt^2 \rbrace \nonumber \\
e^{-2\varphi} &=& (1 - \frac{\mu}{r^2}) 
\end{eqnarray}
The solution describes the formation and subsequent expansion of a
hole at $t = 0$, and differs from Witten's original decay mode \cite{Wit1}
only by the presence of an additional scalar field in four dimensions
(originating from $A_{\tau}$).

\bigskip

As in \S 3, 
we can also consider dimensional reduction along the Killing vector 
$\partial_{\tau} + B \partial_{\phi}$ where $B = n/R$. 
The four-dimensional fields obtained are:
\begin{eqnarray}
ds_{ein}^{2} &=& e^{-\varphi} \lbrace \frac{dr^2}{(1-\frac{\mu}{r^2})} + r^{2}
(d\theta^{2} - \rm{cos}^{2}\theta dt^{2}) 
 + e^{2\varphi}(r^{2} - \mu) \rm{sin}^{2}\theta d\psi^{2} 
\rbrace \nonumber \\
 e^{-2\varphi} &=& \lbrace (1 - \frac{\mu}{r^2}) + (1 -
 \frac{\kappa\mu}{r^2})^{2} B^{2} r^2 \rm{sin}^{2}\theta \rbrace \\
A^{1}_{\phi} &=& e^{2\varphi} (1 - \frac{\kappa\mu}{r^2})^{-2}
 B^2 r^2 \rm{sin}^{2} \theta \nonumber \\
\cal{A} &\equiv & A_{\tau} = - \frac{\rm{sin}\delta}{\sqrt{2}(1-\kappa)}
\frac{r^2-\mu}{r^2-\kappa\mu} \nonumber 
\end{eqnarray}
describing the pair creation of monopoles within a background magnetic
field $A^1$ and background scalar fields $\cal{A}$ and $\varphi$ 
(which are asymptotically constant). However, the magnetic field $B
= n/R$ once again lies outside the range of validity $B
\ll 1/R$, and 
we need to consider rotating black holes to obtain
magnetic fields of physical magnitude. 

\bigskip

For simplicity, we consider a five-dimensional black hole solution
with only one electric charge $Q_f$, and one rotational parameter
$a$ non-zero. Such a solution may be obtained from 
boosting the (Lorentzian) Myers-Perry solution; the most general such
solutions are discussed in \cite{CY}. Rotating $t \rightarrow i\tau$,
$a \rightarrow - i\alpha$ and $Q_f \rightarrow -i Q_f$, 
we obtain the Euclidean section (in the string frame):
\begin{eqnarray}
ds_{str}^{2} &=& \Sigma \lbrace \frac{(\Sigma-\mu)}{\Delta} d\tau^2 
  + \frac{dr^2}{(\rho^2-\alpha^2-\mu)} +
  d\theta^2 + \frac{\rho^2\rm{cos}^2\theta}{\Sigma}d\chi^2 \nonumber \\
&& - \frac{\mu\alpha\rm{sin}^2 \theta}{\Delta}(1+
\rm{cos}\delta)d\tau d\psi + \frac{\rm{sin}^2\theta}{\Delta} 
(\Delta - \alpha^2\rm{sin}^2\theta(\Sigma +
\mu\rm{cos}\delta))d\psi^2  \rbrace \nonumber \\
\Delta &=& (\Sigma - \kappa\mu)^2 \nonumber \\
e^{-2\phi} &=& \frac{\Delta}{\Sigma^2} \\
A_{\tau} &=& - \frac{\rm{sin}\delta}
{\sqrt{2}(1-\kappa)}\frac{\Sigma-\mu}{\Delta^{1/2}} \nonumber \\
A_{\psi} &=& - \frac{\mu\alpha\rm{sin}\delta \rm{sin}^2\theta}
{\sqrt{2}\Delta^{1/2}} \nonumber 
\end{eqnarray}
where $\Sigma = (\rho^{2} - \alpha^{2}\rm{cos}^{2}\theta)$, $\kappa$
is defined as previously and we have included a pure gauge term in $A_{\tau}$
so that $A^2$ is non-singular at the poles of the horizon $\rho_{h}^2 =
\mu + \alpha^2$. The charge 
$Q_{f} = (\sqrt{2}/8)\pi \mu \rm{sin}\delta$ 
using the same conventions as previously. 

To avoid a conical singularity at the horizon, we choose the
radius at infinity to be
$R = (1 - \kappa)/\rho_{h}$; the Euclidean angular
velocity is $\Omega = \alpha/\mu(1- \kappa)$. The action is easily
calculated from (\ref{sa}), with the background subtraction facilitated
by the flatness of the $\mu =0$ solution for all values of $\alpha$; 
then:
\begin{equation}
S_{E} = \frac{\pi\mu (1+ 6\kappa)}{8G_{4}} = \frac{\pi R^2 (1+
  6\kappa)}{8G_{4}(1-\kappa)^2 (1 - \Omega^{2}R^{2})}
\end{equation}
As usual, we rotate a coordinate on the sphere $\chi \rightarrow it$ 
to obtain the subsequent Lorentzian evolution. Then,
dimensional reduction along $l = \partial_{\tau} + \Omega
\partial_{\psi}$ leads to a bubble decay mode, and reduction along 
$l' = \partial_{\tau} + (\Omega + n/R)\partial_{\psi}$ describes
monopole pair production, with the decay rate of the latter suppressed
since the action is greater. For the latter,  
magnetic fields of physical magnitude and
avoidance of conical singularities in the four-dimensional solution 
require $n = \pm 1$ and $\left| \Omega \right| \approx 1/R$. 

\bigskip

We have considered only the most simple charged rotating solution;
the prescription for obtaining the most general decay modes is as follows.
Starting from the most general five-dimensional Lorentzian rotating,
(electrically) charged black hole solution \cite{CY}, we look for a Euclidean
section on which all fields can be chosen to be real. If such a
section exists, then by Witten
rotating a coordinate on the sphere, we can obtain a vacuum decay
mode. Dimensional reduction along a Killing vector with fixed point
set of dimension three leads to decay by bubble formation, a decay
process lying in a different superselection sector of the Hilbert
space to the supersymmetric vacuum, whilst dimensional reduction 
along a Killing vector with
a fixed point set of dimension one leads to decay by monopole pair
production, a decay process consistent with the spin structure of the 
background. Finally, by rotating the torus coordinates (allowing for
non-trivial angles between the generating circles), we obtain the
generalisations of the solutions discussed in \S 3. 

All such decay modes do not describe the decay of the supersymmetric
vacuum, are incomplete at null infinity, and have actions greater than
the action of the original decay mode of Witten described in \S3. 

\section{Extremal black holes as instantons}
\noindent

The discussion in the previous sections has been based around
five-dimensional black hole solutions of topology $R^2 \times S^3$,
whose asymptotic geometry is that of the background $R^4 \times S^1$. 
However, extremal black holes are believed to have the
topology $S^1 \times R \times S^3$, with the Killing vector in the
circle direction having no fixed point sets \cite{GK}. 
In contrast to the choice for 
non-extremal solutions we must choose a spin structure such that
spinors are 
periodic in this direction, and there is hence no obstruction to 
the analytically continued solutions asymptotically admitting the 
covariantly constant spinors of the background. 

\bigskip
 
To illustrate this, we consider a particular five-dimensional extremal
black hole solution to the equations of motion which follow from 
(\ref{Act}), carrying electric charges with respect to both gauge
fields where:
\begin{eqnarray}
Q_{\bar{f}} &=& \frac{1}{4\pi^{2}} \int \ast e^{\frac{4\phi}{3}} \bar{F} \\
Q_{f} &=& \frac{1}{16\pi} \int \ast e^{-\frac{2\phi}{3}} F \nonumber 
\end{eqnarray}
For a spherically symmetric solution we have:
\begin{eqnarray}
\ast e^{\frac{4\phi}{3}} \bar{F} &=& 2 Q_{\bar{f}} \epsilon_{3} \\
\ast e^{-\frac{2\phi}{3}} F &=& \frac{8Q_{f}}{\pi} \epsilon_{3}
\nonumber 
\end{eqnarray}
and there exist solutions with constant dilaton such that:
\begin{equation}
e^{2\phi} = 2 (\frac{\pi Q_{\bar{f}}}{4Q_{f}})^{2}
\end{equation}
The field equations then imply that the metric takes the
Reissner-Nordstrom form:
\begin{eqnarray}
ds_{ein}^{2} &=& - (1 - (\frac{r_{0}}{r})^{2})^{2} dt^2 + 
(1 - (\frac{r_{0}}{r})^{2})^{-2} dr^{2} +
r^{2}d\Omega_{3}^{2} \nonumber  \\
r_{0} &=& (\frac{8 Q_{\bar{f}}Q_{f}^{2}}{\pi^{2}})^{\frac{1}{6}}
\end{eqnarray}
We consider these extremal solutions (with both charges non-zero)
since extremal solutions with only one charge non-zero have degenerate
horizons with zero area, and thus the Euclidean sections have naked
singularities, and cannot be interpreted as an instantons. 
The above solution is the simplest extremal black hole with a
non-degenerate horizon, and for this reason the corresponding
dual solution in IIA theory compactified on $K3 \times T^2$
was recently discussed in the context of the microscopic description
of the entropy \cite{SV}. 

\bigskip

We now attempt to analytically continue the Lorentzian solution into the
Euclidean regime, by rotating $\tau = it$; now, the gauge fields in 
the original solution are:
\begin{eqnarray}
\bar{F} &=& \frac{16}{\pi^{4/3}r^{3}} Q_{f}^{4/3} Q_{\bar{f}}^{-1/3} dt
\wedge dr \nonumber \\
F &=& \frac{8}{\pi^{1/3}r^3} Q_{f}^{1/3} Q_{\bar{f}}^{2/3} dt \wedge dr 
\end{eqnarray}
When we rotate $t \rightarrow i\tau$, if we impose the requirements
that the dilaton field and $r_{0}^{2}$ are real, 
both $Q_{f}$ and $Q_{\bar{f}}$ remain real and positive, 
so that the gauge fields become pure imaginary. 
If however we impose the requirements that the gauge fields
are real on the Euclidean section, then $r_{0}$ becomes complex, and
the metric is not real. Thence a Euclidean section on which all the
fields are purely real does not exist.

If we take (electric) field strengths that are pure imaginary 
on the Euclidean section, our solution takes the form:
\begin{eqnarray}
ds_{ein}^{2} &=& (1 - (\frac{r_{0}}{r})^{2})^{2} d\tau^{2} + 
(1 - (\frac{r_{0}}{r})^{2})^{-2} dr^{2} + r^2 d\Omega_{3}^{2}
\nonumber \\
\bar{F} &=& i \frac{16}{\pi^{4/3}r^{3}}Q_{f}^{4/3} Q_{\bar{f}}^{-1/3} 
d\tau \wedge dr \\
F &=& i \frac{8}{\pi^{1/3}r^{3}} Q_{f}^{1/3} Q_{\bar{f}}^{2/3}
d\tau \wedge dr 
\nonumber 
\end{eqnarray}
As before, we need to include (pure imaginary) gauge terms to the Maxwell
potentials $A$ and $\bar{A}$ to ensure that both $A^2$ and $\bar{A}^2$
are non-singular at the horizon.
There is no naturally defined periodicity of $\tau$ and we define the 
periodicity at spatial infinity as $\beta$; the action 
\footnote{Since the metric is of the Reissner-Nordstrom form, the
  subtleties in calculating the boundary terms considered 
  in \cite{GK} do not arise here.}
of the Euclidean solution (\ref{sa}) is then:
\begin{equation}
S_{E} = \frac{\pi r_{0}^{2} \beta}{8 G_{5}} = \frac{\pi r_{0}^{2}}{8 G_{4}}
\end{equation}
which implies a vanishing entropy in the semi-classical
approach \cite{HR} since:
\begin{equation}
S = (\beta \partial_{\beta} - 1) S_{E} = 0
\end{equation}
(although string theory 
calculations give a non-zero answer and 
it is believed that string corrections as the length of the imaginary time
direction approaches the string scale lead to a non-vanishing entropy 
as well as perhaps changing the topology to that of non-extremal 
solutions \cite{H}). 

For comparable values of the parameters $r_{0}$ and $\mu$, 
the decay rate by this mode is similar to that in \S3; the range of
validity of the semiclassical calculation requires that $r_{0}$ is
much larger than the Planck length, and hence the rate of vacuum decay
is necessarily slow. However, since
$\beta$ is not fixed by the solution, we can choose a radius at
infinity consistent with a Kaluza-Klein interpretation (unlike the
solutions in the previous whose internal directions are too large for
such interpretations). 

As in \cite{Wit1}, we construct a bubble decay mode of the vacuum by 
rotating a coordinate on the sphere. The
mass on the initial value hypersurface vanishes, since the perturbation 
falls off faster than $1/r$, as do the charges of the fields; hence
the solution may describe the decay of the vacuum state. The
decay mode again involves the formation and subsequent
expansion of a bubble of radius $r_0$, with null infinity incomplete. 

Since
there is no obvious obstruction to finding asymptotically constant
spinors in this solution, it may seem at first sight as though the 
instanton represents a possible decay mode of the supersymmetric
vacuum. However, the gauge fields remain imaginary in the Lorentzian
continuation, and hence violate the dominant energy condition on the
energy momentum tensor. Although the total charge is actually zero, 
there are non-vanishing pure imaginary Maxwell potentials.
That is, although the solution has the
requisite asymptotic spin structure for it to contribute to the decay
of the vacuum, it is excluded by the unphysical behaviour of the
energy momentum tensor. 

\bigskip

More generally, {\it{any}} analytically continued extremal solution which  
asymptotically admits the constant spinors of the vacuum on a
hypersurface of zero mass and charge must have an energy momentum
tensor which does not satisfy the dominant energy condition. We justify
this statement in the following section by considering the 
formulation of Witten's proof of the positive energy theorem in 
higher dimensions. 

\section{The positive energy theorem}
\noindent

In the examples that we have discussed so far, the instantons have 
fallen into three categories. Firstly, extremal black hole instantons 
admitting 
isometries with no fixed point sets have spin structures consistent
with the background, but the energy momentum tensors of the
analytically continued solutions do not satisfy the dominant energy
condition (at least in the example we gave). Secondly, in non-extremal
black hole instantons, if we consider
dimensional reduction along a Killing vector admitting an eight
dimensional fixed point set, spinors must be antiperiodic about an
orbit at infinity of this isometry, and hence the solution lies in
a different superselection sector of the Hilbert space to the vacuum.
Thirdly, again in non-extremal black hole instantons, if we consider
dimensional reduction along a Killing vector admitting a
six-dimensional fixed point set, we obtain a solution consistent with
the decay of a magnetic vacuum in four dimensions. 
We now discuss more generally the existence of instantons, and 
show that there are no decay modes consistent both with the
dominant energy condition and with the supersymmetric spin structure 
of the background. 

\bigskip

Yang-Mills instantons do not indicate a possible decay mode of the
vacuum since there does not exist a surface of constant time from
which we can continue them as real Yang-Mills fields in Minkowski
space. The analogue for ten-dimensional supergravity would be a
nine-dimensional surface with zero second fundamental form; such a
surface acts as a ``turning'' point at which the instanton matches the
space into which the vacuum decays. 

In the previous sections, starting from a Lorentzian spacetime we
constructed Euclidean solutions by rotating $(t,x^m) \rightarrow
(i\tau,x^m)$. Even if the original Lorentzian solution did not admit
a hypersurface of constant time of zero momentum we could find a
Euclidean section on which the metric was real, but only by
analytically continuing a momentum parameter in the solution. When we
look for a Lorentzian section of a Euclidean solution, to describe the
subsequent evolution from an initial value hypersurface, we cannot
analytically continue momentum parameters and such a section will only
exist if there is a zero momentum hypersurface. 

Starting from a ten-dimensional Euclidean solution $g_{MN}$ which
admits a nine-dimensional complete hypersurface $\Sigma$ with induced
metric $\hat{g}_{MN}$, we analytically continue back to a Lorentzian  
signature by rotating $(\tau,x^m) \rightarrow (-it,x^m)$. We choose 
the $0$-vector to be orthogonal to $\Sigma$ and, on the 
Lorentzian section, we have a unit timelike normal to $\Sigma$ given by
$t_{M} = g_{0M}/\sqrt{-g_{00}}$ with the induced metric being:
\begin{equation}
\hat{g}_{MN} = g_{MN} - t_{M}t_{N}
\end{equation}
and the second fundamental form of $\Sigma$ is:
\begin{equation}
{\cal K}_{MN} = t_{(P;Q)} \hat{g}^{Q}_{M} \hat{g}^{P}_{N}
\end{equation}
Reality of the metric on the Lorentzian section requires that the
$g_{0m}$ terms vanish, which implies that $\hat{g}_{0m} = {\cal K}_{00} =
{\cal K}_{0m} = 0$ and:
\begin{equation}
{\cal K}_{mn} = \frac{1}{2\sqrt{-g_{00}}} g_{nm,0}
\end{equation}
The surface $\tau = \tau_{0}$ of the Euclidean
solution must match that the surface $t = t_{0} = i \tau_{0}$ of the
analytically continued solution; a necessary condition
is that $g_{nm,0}|_{t_{0}} = 0$. Hence, if an instanton is to describe
the decay of the vacuum, it must admit a surface with zero extrinsic
curvature from which we can analytically continue the solution. We
usually find such a surface by looking for a hypersurface orthogonal
Killing vector in the Euclidean solution and Wick rotating. 

\bigskip

Now, if one has such a hypersurface of constant time, Witten's
proof of the positive energy theorem can be applied unless 
there is some obstruction such as a black hole. 
Suppose that we have an instanton for which the fixed point sets of
the isometries do not prevent us from finding asymptotically constant
spinors; the solution then lies in the same superselection sector as
the vacuum. However, the absence of any obstruction on the
initial value hypersurface allows us to prove the positivity of the
mass by Witten's method; that is, if the solution is not flat, the
mass must be positive, and the vacuum cannot decay into this state.   

Putting it this way makes it sound as though the decay modes of the
supersymmetric vacuum must be trivial; if a black hole type of 
obstruction is present, the positive energy theorem does not apply but
the instanton does not lie in the same superselection sector of
fermions as the vacuum. In the absence of an obstruction the
positive energy theorem applies and prevents the existence of
instantonic decay modes unless the Witten proof fails in another
(unphysical) way such as the dominant energy condition breaking down, 
which we discussed in \S5. We now discuss the formulation of the
positive energy theorem for spacetimes asymptotic to $M^4 \times T^6$.
 
\bigskip

We consider an asymptotically flat solution to the Euclidean field 
equations derived from the Einstein frame action in ten dimensions 
(\ref{einfr}); the graviton field equation gives:
\begin{equation}
R_{MN} - \frac{1}{2} R g_{MN} = 8 \pi G_{10} T_{MN} \label{field}
\end{equation}
where the energy momentum tensor includes contributions from the
dilaton, antisymmetric tensor and gauge fields. The field 
configuration must also be consistent with the other constraint 
equations derived from the action.

We then analytically continue the solution to obtain a Lorentzian
solution, imposing the condition that the energy momentum tensor satisfies 
the dominant energy condition \cite{HE}; that is, the local energy density
$T_{00}$ is positive (or zero) at each point in the Lorentzian
spacetime and in each local Lorentz frame. The total energy momentum
tensor is:
\begin{eqnarray}
T_{MN} &=& \frac{1}{8} \lbrace (\partial_{M}\Phi)(\partial_{N} \Phi) -
\frac{1}{2} (\partial\Phi)^2 g_{MN} \rbrace 
+ \frac{1}{4} e^{-\frac{\Phi}{2}} \lbrace g_{NN'}H_{MPQ}H^{N'PQ}
- \frac{1}{6} H^2 g_{MN} \rbrace \nonumber \\
&& + \frac{1}{2} e^{-\frac{\Phi}{4}} \lbrace g_{NN'}F^{(f)}_{MP}F_{(f)}^{NP} -
\frac{1}{4} Tr(F^2) g_{MN} \rbrace  
\end{eqnarray}
It is well known that each of the
contributions to the energy momentum tensor obey the
required condition, provided that the fields are real. Since this
condition is critical to the proof, we include a short discussion of
the dominant energy condition in the appendix. 

Evidently reality of the fields
is a non-trivial constraint when we consider analytic continuation of
Euclidean solutions. F will be real in Lorentzian spacetime provided
that $F_{\tau m}$ is pure imaginary and $F_{mn}$ is pure
real; that is, the magnetic field must be real and the electric field
imaginary on the Euclidean section. Similarly, the reality of H in the
Lorentzian spacetime is ensured by $H_{\tau mn}$ being pure imaginary, and
$H_{mnp}$ being pure real. 

We choose to work in the Einstein frame since the energy momentum tensor
in the string frame does not satisfy the dominant energy condition
\cite{GM}, \cite{HHSK}; the graviton field equation
obtained from the string frame action (\ref{boselagrangian}) is:
\begin{equation}
R_{MN}(G) - \frac{1}{2} R(G) G_{MN} = 8 \pi G_{10} T^{str}_{MN} 
\end{equation}
where the total energy momentum tensor is defined as:
\begin{eqnarray}
T_{MN}^{str} &=& - 2  \lbrace (\partial_{M}\Phi)(\partial_{N} \Phi) -
\frac{1}{2} (\partial\Phi)^2 G_{MN} \rbrace + \frac{1}{4} \lbrace 
G_{NN'}H_{MPQ}H^{N'PQ} - \frac{1}{6} H^2 G_{MN} \rbrace \nonumber \\
&& + \frac{1}{2} \lbrace G_{NN'}F^{(f)}_{MP}F{(f)}^{N'P} -
\frac{1}{4} Tr(F^2) G_{MN} \rbrace  
\end{eqnarray}
Hence, the dilaton contribution to the energy momentum tensor no longer
satisfies the dominant energy condition, because of the change of sign
of the $(\partial\Phi)^2$ term in the action.

\bigskip 

Since the solution is asymptotically flat, we can decompose the 
metric at spatial infinity as:
\begin{equation}
g_{\hat{M}\hat{N}} = 
\left ( \matrix{- \delta_{\hat{0}\hat{0}} + 
h_{\hat{0}\hat{0}}(x^{p})
& h_{\hat{0}\hat{m}}(x^{p}) \cr
 h_{\hat{n} \hat{0}} (x^{p}) & \delta_{\hat{m}\hat{n}} + 
h_{\hat{m}\hat{n}}(x^{p})} \right) 
\end{equation}
where the $h_{\hat{0}\hat{m}}$ terms vanish if $\Sigma$ has zero momentum.  
We use asymptotically flat coordinates, and work in an orthonormal frame.
If the perturbation to the flat metric is of order $(1/r^k)$; 
derivative terms are of order $1/r^{k}$ for $\partial_i$
terms and of order $1/r^{(k+1)}$ for $\partial_{\mu}$ terms.

For a solution admitting a hypersurface of asymptotic geometry 
$R^3 \times T^6$ the
ADM energy can be expressed as \cite{BKKLS}:
\begin{equation}
E_{ADM} = \frac{1}{16 \pi G_{10}} \oint_{\infty} d\Sigma^{m} 
(\partial_{n} h_{nm} - \partial_{m} h_{nn})
\label{ene2}
\end{equation}
where the integral is taken over a boundary at infinity of $\Sigma$; the ADM
momentum is given by:
\begin{equation}
P_{\hat{m}} = \frac{1}{16\pi G_{10}} \oint_{\infty} d\Sigma^{\hat{n}}
(\partial_{\hat{n}} h_{\hat{0}\hat{m}}-\partial_{\hat{0}} h_{\hat{n}\hat{m}}
 + \delta_{\hat{n}\hat{m}} \partial_{\hat{0}} h_{\hat{p}\hat{p}} - 
 \delta_{\hat{n}\hat{m}} \partial_{\hat{p}} h_{\hat{0}\hat{p}})
\end{equation}
Expanding the expression for the ADM energy in terms of the
torus coordinates $x^{i}$ and the external coordinates
$x^{\alpha}$:
\begin{equation}
E_{ADM} = \frac{1}{16 \pi G_{10}} 
\oint_{\infty} d\Sigma^{\alpha} 
\lbrace \partial_{\beta}h_{\beta\alpha}
- \partial_{\alpha}h_{ii} - \partial_{\alpha}h_{\beta\beta} \rbrace
\label{ene}
\end{equation}
where the $\partial_{i}h_{i\alpha}$ terms vanish when we consider a solution
admitting a global $U(1)^6$ isometry group. 
Since the volume element is of
order $r^2$, the energy is only well-defined when the integrand is of order
$1/r^2$ which requires that $h_{\alpha\beta}, h_{ii}$ are of order $1/r$. 
Since the functions 
$h_{i\alpha}$ must be periodic in the coordinates $x^{i}$, even if they are
not independent of the torus coordinates, 
the expression for the ADM energy in (\ref{ene}) is valid even when the 
metric perturbations are dependent on the torus coordinates. 

\bigskip

To prove the positive mass theorem for solutions which tend to the required 
background we consider spinors obeying a Dirac type equation on the 
nine-dimensional hypersurface. Our discussion follows closely that of
\cite{GHHP}, and although we are interested in zero momentum
hypersurfaces for generality we do not impose the requirement 
that ${\cal K}=0$, since it is not required by the proof. 
Projecting the ten-dimensional covariant 
derivative $D_{M}$ onto the hypersurface:
\begin{equation}
D_{\hat{m}}\epsilon = (\nabla_{\hat{m}} + \frac{1}{2} 
{\cal K}_{\hat{m}\hat{n}} \gamma^{\hat{n}} \gamma^{\hat{0}}) \epsilon 
\end{equation}
where $\nabla$ is the covariant derivative on the hypersurface, the gamma 
matrices are 
constructed from the $32$-dimensional spinor representation of $SO(9,1)$
and $D_{M} = \partial_{M} + \Gamma_{M}$ with $\Gamma_{M}$ the spin connection
matrices. Then, multiplying by $\gamma^{\hat{m}}$, we
obtain the Witten equation:
\begin{equation}
\gamma^{\hat{m}} \nabla_{\hat{m}} \epsilon = 
- \frac{1}{2} {\cal K} \gamma^{\hat{0}} \epsilon \label{wit}
\end{equation}
with ${\cal K}$ the trace of the second fundamental form. 
If we multiply by $\epsilon^{\ast}$, act on the result with
$\gamma^{\hat{m}} \nabla_{\hat{m}}$ and use the Ricci identity, we obtain:
\begin{eqnarray}
\nabla_{m} (\epsilon^{\ast} D^{m} \epsilon) &=& 
(D_{m} \epsilon)^{\ast} (D^{m} \epsilon) 
 + \frac{1}{4} \epsilon^{\ast} \lbrace {\cal{R}} + {\cal K}^{2} -
{\cal K}_{mn} {\cal K}^{mn} \label{po}\\
&& \ + 2 \nabla_{m} ({\cal K}^{mn} - \hat{g}^{mn} {\cal K}) 
\gamma_{n} \gamma^{\hat{0}} \rbrace \epsilon \nonumber 
\end{eqnarray}
where $\cal{R}$ is the Ricci scalar on $\Sigma$ of the induced metric. 
The field equations are:
\begin{eqnarray}
{\cal{R}} + {\cal K}^{2} - {\cal K}_{mn} {\cal K}^{mn} 
&=& 16 \pi G_{10} T_{\hat{0}\hat{0}} \\
\nabla_{m} ({\cal K}^{mn} - h^{mn} {\cal K}) 
&=& 8 \pi G_{10} T^{\hat{0}n} \nonumber
\end{eqnarray}
and since the total stress energy tensor satisfies the 
dominant energy condition implying that  $T_{\hat{0}\hat{0}} 
\geq \left| T_{\hat{m}\hat{n}} \right|$:
\begin{equation}
\nabla_{m} (\epsilon^{\ast}D^{m}\epsilon) \geq (D_{m}\epsilon)^{\ast}
(D^{m}\epsilon)
\end{equation}
Upon integrating this over the initial value hypersurface, we obtain:
\begin{equation}
\oint_{\infty} \epsilon^{\ast} D_{m} \epsilon d\Sigma^{m} - \oint_{H}
\epsilon^{\ast} D_{m} \epsilon d\Sigma^{m} \geq \int_{\Sigma} (D_{m}
\epsilon)^{\ast} (D^{m} \epsilon) d\Sigma
\label{a1}
\end{equation}
where we integrate over the region of $\Sigma$ bounded by an inner
surface $H$ and a surface at infinity. In the following, we shall
assume that the inner surface term vanishes; this is certainly true
if $H$ is an apparent horizon or a minimal surface in a maximal
hypersurface (since the proofs given in
\cite{GHHP} are easily generalised to ten dimensions). This covers all
cases of inner boundaries in which we are interested. 

Let the surface term at infinity be {$\cal{S}$}; since the right-hand side is
positive semi-definite, {$\cal{S}$} is also positive semi-definite and
is an invariant of the initial value hypersurface. The contributing
terms to the integrand are of order $1/r^2$, with the $1/r$ contributions
vanishing as:
\begin{equation}
\oint_{\infty} dx^{i}(\partial_{i}h_{mn}) = \oint_{\infty}
dx^{i}\partial_{i}\epsilon = 0
\end{equation}
even when $h_{mn}$ is $x^{i}$ dependent, since any such dependence
must be periodic in $x^{i}$. In fact, if we assume that there are no
Kaluza-Klein type charges arising from the torus, then
$\partial_{i}h_{mn}$ and $h_{\mu i}$ fall off as $1/r^2$ and the
leading order deviation will be independent of these terms in any case. 

We now demonstrate the relationship between {$\cal S$} and the energy 
by considering solutions to Witten's equation; we omit much
of the analysis since it follows as a direct generalisation
of that in \cite{Wit2}. Now,
no non-zero spinor satisfying Witten's equation on the initial value 
hypersurface vanishes as $r \rightarrow \infty$. This follows directly
from (\ref{a1}) since if we assume that a 
solution to Witten's equation which vanishes at infinity exists, 
the left hand side of (\ref{a1}) vanishes, 
as such a solution must decay at least as fast as $1/r^2$ (for
asymptotic geometry $R^3 \times T^6$).
The energy-momentum tensor then vanishes, and $D_{m}\epsilon 
\equiv 0$; if however $D_{m}\epsilon \equiv 0 $ and $\epsilon \neq 0$,
then $\epsilon$ does not vanish at infinity, thus completing the proof. 

We consider a solution to the Witten equation asymptotically
approaching a constant spinor $\epsilon_{0}$; the asymptotic
geometry constrains $\epsilon - \epsilon_{0} =
\tilde{\epsilon}(\theta,\phi,x^i)/r)$. The 
existence of such solutions is non-trivial
even in the four-dimensional case, and was the subject of \cite{R} and
\cite{PT}. Here, however, such spinors exist by assumption, since we
are only interested in solutions lying in the same superselection sector
of the Hilbert space as the vacuum, requiring that the
constant spinors of the vacuum are asymptotically admitted. 
Since $\epsilon$ is a solution of the Witten equation (\ref{wit}), we
find that:
\begin{equation}
\tilde{\epsilon} = lim_{r \rightarrow \infty} \lbrace r^2 \gamma^{r}
\gamma^{m} \Gamma_{m} \epsilon_{0} + r \gamma^{r}
(\gamma^{\alpha} \partial_{\alpha}) \tilde{\epsilon} \rbrace
\label{spn}
\end{equation}
where we use only the $1/r^2$ terms in $\Gamma_{m}$. Then substituting
$\epsilon$ in the expression for {$\cal S$}:  
\begin{equation}
{\cal{S}} = \oint_{\infty} d\Sigma^{\alpha} 
\lbrace \epsilon_{0}^{\ast} \Gamma_{\alpha} \epsilon_{0} +
\epsilon_{0}^{\ast} \partial_{\alpha} (\frac{\tilde{\epsilon}}{r}) 
\rbrace \label{sz1}
\end{equation}
where we again retain only terms of order $1/r^2$ in the integrand. Using
(\ref{spn}), we find that:
\begin{equation}
\oint_{\infty} d\Sigma^{r} \epsilon_{0}^{\ast} \frac{\tilde{\epsilon}}{r^2} = 
 \oint_{\infty} d\Sigma^{r} \lbrace \epsilon_{0}^{\ast} \gamma^{r} \gamma^{m}
 \Gamma_{m} \epsilon_{0} + \epsilon_{0}^{\ast} \frac{\gamma^{r}}{r^3}
 (\gamma^{\alpha} \partial_{\alpha})\tilde{\epsilon} \rbrace
\end{equation}
In the limit of spatial infinity only the first term contributes to 
(\ref{sz1})and hence:
\begin{equation}
{\cal S} = \oint_{\infty} d\Sigma^{\hat{\alpha}} \epsilon_{0}^{\ast} 
(\Gamma_{\hat{\alpha}} - 
\gamma_{\hat{\alpha}} \gamma^{\hat{n}} \Gamma_{\hat{n}} ) \epsilon_{0}
\end{equation}
where we retain only terms of order $1/r^2$ in $\Gamma_{\hat{m}}$. 
Now the linearised spin connection matrices are defined by:
\begin{equation}
\Gamma_{m} = \frac{1}{16} \lbrace
\partial_{N}h_{Mm} - \partial_{M} h_{Nm} \rbrace [\gamma^{M},\gamma^{N}] 
\end{equation}
and hence substituting for $\Gamma$
and comparing with the expressions for the ADM energy and
momentum, we obtain the relationship:
\begin{equation}
{\cal S} = 4 \pi G_{10} \epsilon^{\ast}_{0} 
(E_{ADM} + P_{\hat{m}}\gamma^{\hat{m}} \gamma^{\hat{0}} ) \epsilon_{0} 
\end{equation}
The non-negativity of {$\cal S$} then implies that 
$E_{ADM} \geq \left|{P^{\hat{m}} }\right|$;
thence the mass is positive semi-definite, vanishing if and only if 
$D_{m} \epsilon = 0$ and the energy momentum tensor is null. The integrability
condition on the spinors and 
the existence of a basis of such covariantly constant spinors
then imply that $R_{mnMN}=0$, with Einstein's equations leading to $T_{MN}=0$
and thus $R_{MNPQ}=0$. That is, the mass vanishes if and only if the spacetime
is flat, and there is no state with the requisite spin structure into
which the vacuum can decay.
As usual, by replacing the covariant derivative with a modified
derivative dependent on the gauge fields, Bogomolnyi bounds
relating the mass and charges may be derived. 

\bigskip

We next discuss the ways in which the solutions given in the previous
sections ``evade'' the positive energy theorem.
Suppose that we consider a metric on the hypersurface with 
a leading order perturbation of $1/r^{k}$ from the flat metric with 
$k > 1$; then the energy (and momentum) defined in (\ref{ene}) 
necessarily vanishes. If massless Witten spinors are admitted on the 
hypersurface, and the energy momentum tensor satisfies the dominant energy
condition, (\ref{po}) must be satisfied and thus:
\begin{equation}
\int_{\Sigma} \lbrace (D^{m} \epsilon^{\ast})(D_{m} \epsilon) 
+ 4\pi G_{10} \epsilon^{\ast} (T_{\hat{0}\hat{0}} + T^{\hat{0}m}
\gamma_{m} \gamma^{\hat{0}}) \epsilon \rbrace d\Sigma = 0 
\end{equation}
This integral over the hypersurface will vanish if and only if 
$D_{m} \epsilon \equiv 0$ and the energy momentum tensor is null,
which implies that the spacetime is flat. Hence, if we have a metric on the
hypersurface which approaches the background faster than $1/r$, and the 
dominant energy condition is satisfied by the (non-flat) solution, 
we will not be able to find asymptotically constant solutions to
Witten's equation; the topology
change implicit in these perturbations implies an obstruction preventing
the existence of such spinors. 

The five-dimensional black hole solutions considered in \S3 and \S4
evade the positive energy theorem in this way; the masses of the 
analytically continued solutions vanish, but there exist no 
asymptotically constant spinors on the initial value hypersurfaces. 

\bigskip

We are now able to justify the statement in \S5 that
the positive energy theorem implies that there exists no Euclidean section
of certain extreme black hole solutions on which all fields are
real. Suppose we have an analytically continued solution admitting a
hypersurface whose mass and charges vanish, as the fields decay at infinity
as $1/r^2$. The horizon in the original solution must be
non-singular, so that this hypersurface is non-singular. There is
no topological obstruction to finding solutions to the Witten equation on 
the initial value hypersurface, and we can show that:
\begin{equation}
\oint_{\infty} (\epsilon^{\ast} D_{m} \epsilon) d\Sigma^{m} 
- \oint_{H} (\epsilon^{\ast} D_{m} \epsilon) d\Sigma^{m} = 0 \label{con}
\end{equation}
where we must take account of the inner boundary $H$. Following \cite{GHHP},
we choose $\epsilon$ to satsify the constraint equation
$\gamma^{\hat{0}} \gamma^{\hat{1}}  \epsilon = \epsilon$ (with the one
direction normal to $H$) and thus restrict the frredom of $\epsilon$
on $H$ by half, as required. We may
then show that the only contributing inner boundary term is:
\begin{equation}
- \oint_{H} (\epsilon^{\ast} D_{m} \epsilon) d\Sigma^{m}
= - 2 \oint_{H} (\epsilon^{\ast}\epsilon) {\cal J} d\Sigma_{h}
\end{equation}
where ${\cal J}$ is the trace of the second fundamental form of $H$
embedded in the hypersurface $\Sigma$. Now for any such extremal 
black hole solution the second fundamental form of $H$ in the
hypersurface vanishes, and hence the inner boundary term must vanish.

However, equation (\ref{con}) contradicts (\ref{a1}), whose right hand side is
positive definite for a non-flat solution. Since the derivation assumed 
that the energy momentum tensor satisfies the dominant energy
condition we conclude that this condition must break down, and hence
the gauge field strengths must be imaginary in this Lorentzian 
continuation. That is, 
although the solution has the requisite asymptotic spin structure for it to
contribute to the decay of the vacuum, it is excluded by the non physical
behaviour of the energy momentum tensor. Hence, if an extremal black hole
solution with non-degenerate horizon admits a Euclidean section whose
topology is consistent with asymptotically constant spinors, it is
impossible to find an analytic continuation on which all fields are
real. 

\bigskip

Another class of solutions which evade the positive energy theorem are
those obtained from taking the product of Euclidean black p-branes 
of topology $R^{2+p} \times S^{2}$ with a flat time direction
(and flat circle directions) \cite{Do2}. The effective four-dimensional
solutions describe either the a static bubble or a static
monopole pair within background fields and are excluded from
the proofs of the positive energy theorem by spin structure
arguments. It is however easy to show that the resultant
configurations are unstable (because of the negative modes of the
Euclidean p-brane solutions)and in any case such decay modes are
excluded by their spin structures.  

\bigskip

Therefore, to contribute to the decay
of the vacuum, a Euclidean instanton must have a section on which the
metric may be analytically continued to the Lorentzian regime. A
necessary and sufficient condition is the existence of a hypersurface
of zero second fundamental form. If the hypersurface does not admit
the same number of asymptotically constant spinors as the
vacuum, it will not describe the decay of the supersymmetric
vacuum. If the hypersurface does admit asymptotically constant
spinors, they must be covariantly constant in order that the mass is
zero. If we have the required number of constant spinors, the
hypersurface must be precisely $R^3 \times T^6$.  

There are hence no instantons contributing to the decay rate of the
supersymmetric vacuum. {\it{Bubble}} decay modes are inconsistent
with the background spin structure; {\it{monopole}} decay modes
describe the decay of an (unphysical) magnetic vacuum and {\it{extremal black
hole}} instantons are inconsistent with the dominant energy condition. 

\section{Calabi-Yau compactification}
\noindent

The discussions in the previous sections applied to toroidal
compactifications of the heterotic string theory;
compactification on a Calabi-Yau space gives
rise to a theory with (more realistically) $N=1$ supersymmetry 
in four dimensions. The spectrum
of the theory is certainly stable since there are no modes with imaginary 
frequencies \cite{Can} and 
the semi-classical stability of the vacuum will be determined by whether
instanton solutions to the classical field equations exist.

We may immediately exclude the possibility that the Calabi-Yau vacuum can
decay into another supersymmetric state; the supersymmetry will require 
that on an initial value hypersurface two covariantly constant spinors
are admitted on the compact space and four are admitted on the external
space. If however such spinors are admitted globally throughout the 
hypersurface, the hypersurface must be precisely $R^3 \times
K_{SU(3)}$, and the
spacetime must be the vacuum. One might imagine that there exist
instantons corresponding to the tunnelling between different $K_{i}$;
however, cobordism theory requires that the $K_{i}$ must have the same
characteristic Stiefel-Whitney and Chern numbers \cite{M} and thence
are topologically indistinct. 
Thus vacuum instability can again result only from the consideration of 
non-supersymmetric states which are not simple metric products but rather 
contain topological defects. 

\bigskip

An instanton will only describe a possible decay mode of the vacuum if
it admits a nine-dimensional hypersurface of zero second fundamental
form whose geometry asymptotically approaches that of $R^3 \times K_{SU(3)}$.
At infinity one would want there to be two asymptotically covariantly
constant spinors on the Calabi-Yau and four on the $R^3$ as for the
vacuum. The most general decay modes may have non-vanishing
antisymmetric tensor and dilaton fields which are asymptotically
constant; the energy momentum tensor again satisfies the dominant
energy condition provided that the fields are real on the Lorentzian
section. 

Now, the generalised expression for the ADM energy of a solution $g_{MN}$
with respect to a background solution ${}^{\circ}g_{MN}$ is \cite{BKKLS}:
\begin{equation}
E_{ADM} = \frac{1}{16 \pi G_{10}} \oint_{\infty} d\Sigma^{m} \lbrace
{}^{\circ}D_{n} g_{mp}
- {}^{\circ}D_{m} g_{np} \rbrace {}^{\circ}g^{np} 
\end{equation}
where ${}^{\circ}D_{m}$ is the covariant derivative in the
background. We assume that $g_{MN}$ is an analytically continued 
solution to (\ref{field}), with the energy momentum tensor satisfying the 
dominant energy condition, and the field configuration consistent with 
the other constraint equations. We decompose the metric at spatial
infinity as:
\begin{equation}
g_{MN} = 
\left ( \matrix{{}^{\circ}g_{00} + h_{00}(x^{p}) & h_{0m}(x^{p}) \cr
 h_{n0} (x^{p}) & {}^{\circ}g_{mn} + h_{mn}(x^{p})} \right) 
\end{equation}
with $h_{MN}$ decaying as $1/r$ and $h_{0m}$ terms vanishing for a
zero momentum hypersurface. Then we can rewrite the ADM energy as:
\begin{equation}
E_{ADM} = \frac{1}{16\pi G_{10}} \oint_{\infty} d\Sigma^{\alpha} \lbrace
\partial_{\beta}h_{\alpha\beta} - \partial_{\alpha}h_{\beta\beta} 
+ {}^{\circ}g^{ij} ({}^{\circ}D_{i}h_{\alpha j}) 
- {}^{\circ}g^{ij} \partial_{\alpha}h_{ij} \rbrace
\end{equation}
We consider solutions to the Witten equation (\ref{wit}) on the
hypersurface now approaching {\it covariantly} constant spinors; that is we
look for a solution $\epsilon$ approaching $\epsilon_{0}$ where:
\begin{equation}
{}^{\circ}D_{m}\epsilon_{0} = 0 
\end{equation}
and $\epsilon$ approaches $\epsilon_0$ as $1/r^2$.
For such a solution, it follows from the dominant energy condition
that:
\begin{equation}
\oint_{\infty} \epsilon^{\ast} D_{m} \epsilon \geq \int_{\Sigma} 
(D_{m}\epsilon)^{\ast}(D^{m}\epsilon)
\end{equation}
where we assume the only boundary is at infinity.  
Since $\epsilon$ satisfies the Witten equation, it is
straightforward to show that the only contributing terms to 
the invariant {$\cal S$} give: 
\begin{equation}
{\cal S} = \oint_{\infty} d\Sigma^{\alpha} \epsilon_{0}^{\ast} 
(\Gamma_{\alpha} - \gamma_{\alpha} \gamma^{m} \Gamma_{m}) 
\epsilon_{0} \label{m1}
\end{equation}
where we linearise the covariant derivative about the background 
as $D_{m} = {}^{\circ}D_{m} + \Gamma_{m}$, and:
\begin{equation}
\Gamma_{m} = \frac{1}{16} 
({}^{\circ}D_{N}h_{mM} - {}^{\circ}D_{m}H_{MN})
[\gamma^{M},\gamma^{N}] 
\end{equation}
Substituting into (\ref{m1}), and following the same steps as in
\cite{Wit2}, we find that:
\begin{eqnarray}
{\cal S} &=& \oint_{\infty} d\Sigma^{\alpha} \epsilon^{\ast}_{0}\epsilon_{0}
\lbrace ({}^{\circ}D_{m}h_{\alpha n})
{}^{\circ}g^{mn} - ({}^{\circ}D_{\alpha}h_{mn} ){}^{\circ}g^{mn}
\rbrace \nonumber \\
&=& 16 \pi G_{10} \epsilon^{\ast}_{0} \epsilon_{0} E_{ADM}
\end{eqnarray}
where we assume that the hypersurface has zero second fundamental form.
Since {$\cal S$} is positive semi-definite, the ADM energy of any solution
asymptotically approaching this background is also
constrained to be positive semi-definite with respect to the
background. Vanishing of the energy requires that the energy momentum 
tensor vanishes, and $D_{m} \epsilon_{0} = 0$. Since the background
admits a basis of covariantly constant spinors, and $\epsilon_{0}$ is
an arbitrary element of the basis, zero energy requires the existence
of the full number of covariantly constant spinors on the
hypersurface, forcing the hypersurface to be precisely 
$R^3 \times K_{SU(3)}$. 

Even if we assume that the perturbations fall away
sufficiently quickly that the energy vanishes, then the vanishing of
{$\cal S$} implies that the requisite solutions to Witten's equation
(\ref{wit}) are only admitted if 
$D_{m} \epsilon \equiv 0$, again implying that the solution is
precisely the background. Even if other solutions of zero energy 
with respect to the background exist, they cannot have the required 
asymptotic spin structure. Obstructions on the hypersurface may allow the 
evasion of the positive energy theorem, but also imply either an incompatible
spin structure at infinity or a violation of the dominant energy condition. 
Since the instanton cannot contribute to
the decay rate unless one has asymptotically the requisite spinors, 
we conclude that there
can be no instantonic decay modes of the Calabi-Yau vacuum. 
It is similarly straightforward to prove the absence of decay modes of
the $K3 \times T^2$ vacuum using the basis of $16$ covariantly
constant spinors.

\section{Eleven-dimensional supergravity}
\noindent

Our discussions so far have been restricted to ten dimensional
supergravity but the same arguments can be applied to 
eleven-dimensional supergravity.
The bosonic sector of the action of $N=1$ supergravity in 
eleven dimensions is given by \cite{CJS}:
\begin{equation}
S = \int d^{11}x {\cal E} \lbrace \frac{1}{4} R({\cal E}) 
- \frac{1}{48} {\cal F}^{2} + \frac{2}{(12)^{4}}
\epsilon^{x_{1}...x_{11}}{\cal F}_{x_{1}...x_{4}}{\cal
  F}_{x_{5}...x_{8}} {\cal A}_{x_{9}...x_{11}} \rbrace \label{elev}
\end{equation}
where ${\cal E}$ is the elfbein, and ${\cal F}_{wxyz} = 4
\partial_{[w} {\cal A}_{xyz]}$. The conjectured duality between 
eleven dimensional supergravity compactified on a circle
and strongly coupled type IIA superstring theory \cite{WT}
(as well as other
related dualities) suggests that the absence of instantonic decay modes
of the former should for consistency imply the absence of such decays
for the latter. 

The equations of motion derived from
(\ref{elev}) (assuming that the gravitino vanishes) are:
\begin{eqnarray}
R_{yz} - \frac{1}{2}R g_{yz} &=& \frac{1}{3} \lbrace {\cal
  F}_{yx_{1}x_{2}x_{3}}{\cal F}_{z}^{x_{1}x_{2}x_{3}} - \frac{1}{8}
 g_{yz} {\cal F}^{2} \rbrace \nonumber \\
D_{y} {\cal F}^{y x_{1}x_{2}x_{3}} &=& - \frac{1}{576}
\epsilon^{x_{1}...x_{11}} {\cal F}_{x_{4}...x_{7}}{\cal
  F}_{x_{8}...x_{11}}
\end{eqnarray}
Now, if we require a vacuum state in which the three form vanishes, 
the solution must admit at least one covariantly constant spinor 
in order to preserve some supersymmetry. In the context of
compactification to four dimensions, this implies that the holonomy
of the compact space must be a subgroup of $Spin(7)$ \cite{Du}. We
usually take the compact space to be $T^7$, $K3 \times T^3$, 
$K_{SU(3)} \times S^1$ or $K_{Spin(7)}$ 
for which the eleven dimensional vacuum admits
$32$, $16$, $8$ or $4$ covariantly constant spinors respectively. 

We consider the existence of instantons
by, as usual, looking for solutions to the Euclidean equations of
motion whose asymptotic geometry is that of the background. The most
general such solution may have a non-zero three form field which is
asymptotically constant, and consistent with the field equations.

Such
a solution only contributes to the vacuum decay rate if it both admits an
initial value hypersurface from which we can describe the subsequent
Lorentzian evolution and also asymptotically admits the requisite
number of covariantly
constant spinors of the vacuum on this hypersurface. In addition, we
require that the analytic continuation of the four form field strength
${\cal F}$ is real; that is, on the Euclidean section, the
``electric'' part of ${\cal F}$ is imaginary and the ``magnetic'' part
is real. Then,
the energy momentum tensor satisfies the dominant energy condition
(see the appendix) and the spinorial proofs of the
positive energy theorem may be applied. 

However, if all of these
requirements are satisfied, the methods of \S 6 can be used to
show that the energy of the
solution with respect to the background is positive semi-definite and
only vanishes if the solution is identical to the background. That is,
there are no states into which the vacuum can decay, consistent with
asymptotically admitting the covariantly constant spinors of the vacuum.
For compactifications admitting at least one circle factor, we will of
course be able to find bubble and monopole decay modes of a
non-supersymmetric vacuum. 

\section{Conclusions}
\noindent

We conclude by recapitulating the arguments by which we exclude decay
modes of a supersymmetric vacuum solution of supergravity theories in
ten and eleven dimensions. To contribute to the decay of the vacuum, a
Euclidean instanton must have a section on which the metric can be
analytically continued to the Lorentzian regime. A necessary and
sufficient condition is the existence of a hypersurface of zero second
fundamental form. If this hypersurface does not admit the same number
of asymptotically constant spinors of the vacuum, it will not describe
the decay of the supersymmetric vacuum. If the hypersurface does admit
asymptotically constant spinors, they must be covariantly constant in
order that the mass is zero (unless we allow the energy momentum
tensor to be unphysical and violate the dominant energy condition). If
we have the required number of covariantly constant spinors, the
hypersurface must be precisely the vacuum state. Thence, there is no
state into which the vacuum can decay. 

All instantons ``evade'' the positive energy theorem either by
violating the dominant energy condition, or by having an incompatible
spin structure, or by describing the decay of a four-dimensional
magnetic vacuum. The
latter decay modes are consistent with fermions, but involve
unphysical fields, and unconventional identifications on the internal
torus. In addition the sizes of the
internal directions are fixed by the solutions and are necessarily 
too large for a clear Kaluza-Klein interpretation. 

\bigskip

Although we have restricted our discussions to the heterotic string
theory and elven-dimensional supergravity, similar arguments 
apply to the low energy effective actions of
the other string theories. We can exclude, for example, supersymmetric
decay modes of type II theory compactified 
on a Calabi-Yau manifold, $T^2 \times K3$ or
a torus by the analysis of \S6 and \S7.

As an aside, it is interesting to consider the implications of
the dualities between 
theories; the heterotic string compactified to on $T^4$ is dual
to the IIA string on $K3$, and the toroidal vacuum admits a decay mode
by metric field charged brane formation whilst the $K3$ vacuum admits no
decay modes. Now 
the duality relates the heterotic and IIA six dimensional fields
via \cite{Sen2}:
\begin{eqnarray}
\Phi^{h} = - \Phi^{II}, \ \ &&
G^{h}_{AB} = e^{-\Phi^{h}}G^{II}_{AB}, \nonumber \\
A^{h}_{A} = A^{II}_{A},  \ \ &&
M^{h} = M^{II}, \\ 
H^{h} &=& e^{-\Phi} \tilde{H}^{II}, \nonumber
\end{eqnarray}
where the heterotic metric in the string frame is
$G^{h}_{AB}$ and the IIA metric in the string frame is
$G^{II}_{AB}$; $\Phi$, $A$, $H$ are the six dimensional 
dilaton, 24 abelian gauge fields and antisymmetric tensor field
strength respectively. $\tilde{H}$ is the (conformally invariant) dual
tensor to $H$ and 
the $M$ fields are the matrix valued scalar
field representing elements of $O(4,20)/(O(4) \times O(20))$. 
Then, the solution in type II theory dual to that in the heterotic
theory is:
\begin{eqnarray}
ds_{str}^2 &=& e^{\Phi^{II}} \lbrace \frac{dr^2}{(1-\frac{\mu}{r^2})} +
dx^2 + dy^2 + r^2d\theta^2 \nonumber \\
&& -r^2 \rm{cos}^2\theta dt^2 + e^{-2\Phi^{II}} (r^2 - \mu) d\bar{\psi}^2
\rbrace \\ 
e^{2\Phi^{II}} &=& (1 - \frac{\mu}{r^2} + B^2 r^2 \rm{sin}^{2}\theta)
\end{eqnarray}
which corresponds to a solution which is not asymptotically flat in
ten dimensions, and is singular at the ``horizon'' $r = \sqrt{\mu}$.
That is, the dualised solution does not represent an instantonic decay
mode of even a non-supersymmetric vacuum of the dual theory, as we 
would expect.  

\bigskip

Instabilities of vacua of supergravity theories
of the type $M^4 \times K$ exist only if we
include instanton solutions in which the topology is changed.
Even in general relativity,
the necessity of including varying topologies is not obvious, since
cluster decomposition cannot be used to prove the hypothesis (unlike
in Yang-Mills theory). 

It has been suggested that target space duality
in string theory may be used to exclude solutions of different 
topology \cite{CGR}. 
In the string theory, the winding numbers about each of the
compactified directions are conserved quantum numbers. As the radii of
the compactified directions increase, each of the winding numbers
disappears and it is impossible to change the global space topology. 
Since T-duality implies an equivalence relation
between small and large radii of compactified directions, this
suggests that decompactification instability of compactified
directions is not possible either, and thus that no semiclassical
instability of the vacuum exists.
However, the arguments given do not apply to the
Calabi-Yau vacua, since their duality groups do not generally relate small
and large volume compactifications. We might expect that 
topology change should in any case be considered in a wider context 
than string (perturbation) theory. 

In considering the semi-classical stability of the vacua of string 
theories, we have shown that instantons of different topology to
the putative vacuum are excluded by the incompatibility of 
their asymptotic spin structure with that of the vacuum. 
Any instantonic decay modes must lie in a
different superselection sector of the Hilbert space of states, and
do not contribute to the decay rate of the supersymmetric vacuum. 
Although one would expect that the structure of the supersymmetry
algebra at infinity would prevent the existence of even 
non-supersymmetric decay modes, it is reassuring that there is a natural
way of excluding such instantons by semi-classical arguments.

\section{Acknowledgements}

I wish to thank Stephen Hawking for suggesting the 
problem and for useful discussions. I would also like to thank
Gary Gibbons and Paul Townsend for helpful comments. 
Financial support for this work was provided by Trinity College, 
Cambridge. 

\setcounter{section}{0}
\setcounter{equation}{0}
\renewcommand{\thesection}{Appendix \Alph{section}}
\renewcommand{\theequation}{\Alph{section}.\arabic{equation}}

\section{Dominant energy condition}
\noindent

A generic energy momentum tensor $T_{MN}$ satisfies the dominant
energy condition provided that for any future directed timelike or null
$v$ the associated energy momentum flux $j = - T \cdot v$ satisfies
the inequalities:
\begin{equation}
j^2 \leq 0 \ {\rm and } \  -v \cdot j \leq 0  \label{ineq}
\end{equation}
Now, for a $p$-form field ${\cal B}$ with associated 
$(p+1)$-form field strength ${\cal H}$, the energy momentum tensor
is:
\begin{equation}
T_{MN} = \lbrace g_{NN'}{\cal H}_{MP_{1}..P_{p}}{\cal H}^{N'P_{1}..P_{p}}
- \frac{1}{2(1+p)} {\cal H}^2 g_{MN} \rbrace \label{gen}
\end{equation}
It is a standard algebraic exercise to show that such an
energy momentum tensor satisfies
the required condition, assuming that the fields are real. We refer
the reader to \cite{C} and \cite{HHSK} for further details. 

Furthermore, (\ref{ineq}) hold as strict inequalities except in
special circumstances, when either $v$ or the field strengths are
null; that is, the only conditions under which they hold as strict
equalities are as follows:
\begin{eqnarray}
  j^2 = 0 &\leftrightarrow & {\cal H}^2 = 0 \ {\rm or} \  v^2 = 0 \\
- j \cdot v = 0 &\leftrightarrow & j \wedge v = 0 \ {\rm and} \ v^2 = 0
\nonumber \end{eqnarray}
with the latter condition implying that $v$ must be a principle null
vector of ${\cal H}$.

If we assume that the components of ${\cal H}$ are real and that 
${\cal H}^2 < 0$, there exists a frame in which only the components
${\cal H}_{0P_{1}..P_{p}}$ are non zero, ie. the field is purely 
``electric''. It is then straightforward to show that $j^2 \propto
{\cal H}^{4} v^2$ and $-j \cdot v \propto - {\cal H}^{2} v^2$,
implying $j$ satisfies the dominant energy condition. However, for a
purely imaginary ``electric'' field, ${\cal H}^2 > 0$ and the energy 
momentum flux vector does not satisfy the dominant energy condition.
This is a general statement; ${\cal H}$ will satisfy the dominant 
energy condition only if the components are real.  

Since the energy momentum tensors considered in \S6 and \S8 are of
the form (\ref{gen}) (with positive definite conformal prefactors), 
the dominant energy condition is satisfied provided that the fields 
are real on the Lorentzian section.

\end{document}